\documentclass[aps,groupedaddress,amsmath,superscriptaddress,amssymb,prb,twocolumn,nofootinbib]{revtex4-2}
\usepackage{bm}
\usepackage{bbm}
\usepackage{enumerate}
\usepackage{graphicx}
\usepackage[dvips]{epsfig}
\usepackage{epsf}
\usepackage{physics}
\usepackage{cancel}
\usepackage{xcolor}
\usepackage{bbold}
\usepackage{braket}
\usepackage[normalem]{ulem}
\usepackage[caption=false]{subfig}
\usepackage{url}
\usepackage[colorlinks = true,
linkcolor = blue,
urlcolor = blue,
citecolor = blue,
anchorcolor = blue]{hyperref}

\makeatletter
\newcommand{\Rmnum}[1]{\expandafter\@slowromancap\romannumeral #1@}
\makeatletter

\def\tr{\mathrm{tr}}

\def\ep{\varepsilon}
\def\un{^{(n)}}
\def\kb{{\bm k}}

\def\rb{{\bm r}}
\def\ub{{\bm u}}

\def\ka{\varkappa}
\def\0{^{(0)}}
\def\1{^{(1)}}
\def\2{^{(2)}}
\def\3{^{(3)}}
\def\4{^{(4)}}
\def\nh{n_\mathrm{h}}

\def\uh{^\mathrm{(h)}}

\def\SS{\mathbb{S}}

\allowdisplaybreaks

\begin{document}

\title{Effect of hole-strain coupling on the eigenmodes of semiconductor-based nanomechanical systems}

\author{Ankang Liu}
\affiliation{Department of Physics and Astronomy, Michigan State University, East Lansing, Michigan 48824, USA}

\author{Mark Dykman}
\affiliation{Department of Physics and Astronomy, Michigan State University, East Lansing, Michigan 48824, USA}

\begin{abstract}

Electron-phonon coupling can strongly affect the eigenmodes of nano- and micromechanical resonators. We study the effect of the coupling for $p$-doped semiconductor resonators. 
We show that the backaction from the strain-induced redistribution of the holes between and within the energy bands can lead to a nonmonotonic dependence of the modes' eigenfrequencies on temperature and to a strong mode nonlinearity that also nonmonotonically depends on temperature. Unexpectedly, we find that the nonlinearity can nonmonotonically depend on the hole density. We also briefly discuss the effect of the coupling to holes on the modes' decay rates. The results are compared with the experiment.

\end{abstract}

\maketitle

\section{Introduction}

Doping with donors and acceptors is a conventional way to control bulk properties of semiconductors. Besides changing the electron transport, doping also affects vibrational properties, primarily due to the electron-phonon coupling. In the pioneering papers \cite{Keyes1961,*Keyes1968}, Keyes showed that the coupling leads to a significant change in the elasticity parameters of Germanium and Silicon. The effect is ultimately related to the strain-induced breaking of the crystal's cubic symmetry, which leads to changes in the electron and hole band structures and, consequently, the free energy of the electron and hole systems. The change in the elasticity parameters can be thought of as the backaction from the electrons and holes. In the previous theoretical studies the analysis was focused on bulk crystals, and for acceptor-doped crystals \cite{Keyes1961,Bir1963,Cerdeira1972,Kim1976,Averkiev1984,Khan1985} the temperature and acceptor density dependence of the sound velocity was considered. The dependence of the sound velocity on pressure, which is determined by the cubic vibrational nonlinearity, was addressed as well \cite{Averkiev1987}.

New developments occurred when semiconductor crystals began to be used as micro- and nano-scale mechanical resonators \cite{Cleland1996,Cleland1998}; see \cite{Beek2011,Yamaguchi2017,Miller2018,Bachtold2022a} for reviews. Here, doping leads to a change in the eigenfrequencies of the vibrational modes and the temperature and density dependence of these eigenfrequencies \cite{Ng2013}. Moreover, it also affects the nonlinearity-induced dependence of the eigenfrequency on the vibration amplitude \cite{Shahmohammadi2013,Yang2016,Khazaeili2020}. 

Understanding these effects and developing means for describing them is of significant interest. It is also important for numerous applications of nano- and micromechanical resonators that rely on the eigenfrequency remaining almost constant in a broad range of vibration amplitudes and temperatures. 

In this paper, we study the effect of hole-strain coupling on the eigenmodes of micro- or nanoelectromechanical systems (MEMS or NEMS) based on cubic symmetry semiconductors. We consider $p$-doped crystals where, in the energy range of interest, the valence bands consist of light and heavy holes bands and a split-off band, as in Si, Ge, and A$_3$B$_5$ semiconductors. A major effect of strain is the redistribution of holes between the intraband states and between the bands. It can be described analytically, with the account taken of band warping, if the hole density is not too high, so that the split-off band remains empty. The results give the density-dependent hole contribution to the temperature dependence of the mode eigenfrequencies. They also provide the dependence of the eigenfrequencies on the vibration amplitude. The analysis of the full three-band model can be done numerically, and here, besides the theory of the underlying nonlinearity, we extend and modify the numerical results of \cite{Khan1985} to describe the temperature and density dependencies of the eigenfrequencies of several modes usually studied in the experiment. 

The analysis of linear and nonlinear effects of hole-strain coupling on the eigenmodes differs significantly from that of electron-strain coupling in multi-valley semiconductors like Si and Ge \cite{moskovtsev2017strong}. There, the major effect came from the strain-induced lifting of the valley degeneracy and the resulting redistribution of the electrons between the valleys \cite{Keyes1961}; it turned out, however, that, for Si, another major contribution came from the effect of strain on the band crossing \cite{Bir1974,Hensel1965,Laude1971}. The problem of the hole-strain coupling is more complicated, in some sense, as it involves taking into account the strongly nonparabolic dispersion law of the holes. We develop a fairly general formulation that applies both where the problem can be addressed analytically or a numerical analysis is required.

Another important effect of doping and the hole-strain coupling is the change of the mode decay rates. The frequencies of the typically studied eigenmodes are relatively low, $\lesssim 10^{9}$~Hz. An important decay mechanism of such modes is the Akhiezer damping \cite{Akhiezer1938}, cf. \cite{Iyer2016,Rodriguez2019,Bachtold2022a}. It is controlled by thermal phonons. The change of the lifetime of thermal phonons due to doping directly affects the decay rate of the eigenmodes. The full analysis of the effect requires a detailed study of the phonon dispersion law and scattering rates as well as the nonlinear coupling of thermal phonons to each other and to the low-frequency eigenmodes. We use a simple phenomenological model of the Akhiezer damping to obtain an estimate of the effect and to reveal possible consequences of doping on the mode decay.

In Section~\ref{sec:general} we relate the problem of the coupling-induced change of the eigenfrequency and the nonlinearity of a mode of a micro/nanomechanical resonator to the change of the energy and the free energy of the holes. Section~\ref{sec:T_dependence} provides analytical expressions for the coupling-induced change of the linear elasticity tensor within the two-band approximation of the hole dispersion law, with the account taken of the band warping. It also provides numerical results for the temperature dependence of the mode eigenfrequencies for $p$-doped GaAs resonators. In Section~\ref{sec:nonlinearity} we consider the doping-induced correction to the dependence of the vibration frequency on the amplitude in the two-band approximation and provide numerical results on the temperature dependence of the nonlinearity for GaAs-based resonators. Section~\ref{sec:spin_split_off_band} provides the results for the three-band model. In this section, we describe an efficient numerical algorithm that allows us to find the effect of the coupling on the parameters of an eigenmode and present numerical results on the temperature dependence of several eigenmodes of Si-based resonators and the temperature and hole density dependence of the mode nonlinearity. Section~\ref{sec:Akhiezer} gives estimates of the direct hole-induced decay rate of eigenmodes and discusses possible effects of the coupling on their Akhiezer damping rates. In Section~\ref{sec:conclusions} we summarize the results and highlight the implications of this paper. Appendices~\ref{sec:free_energy}--\ref{sec:c44_T_depen} provide details of the calculations.

\section{Coupling of holes to low-frequency vibrational modes}
\label{sec:general}

The typical size of crystalline plates or beams used as nano- and micromechanical resonators ranges from submicron to several hundred microns. Therefore the vibration periods of the low-lying eigenmodes are in the range $10^{-9}$--$10^{-6}$~s. They are much longer than the reciprocal Maxwell relaxation time and the hole thermalization time, which are $\lesssim 10^{-11}$~s for hole densities $\gtrsim 10^{16}$~cm$^{-3}$ and room temperatures. The strain from eigenmode vibrations can be thus assumed quasi-static in the analysis of the hole dynamics. Then the holes in different bands have the same strain-modulated chemical potential $\mu$ and the same temperature. They are described by the density of the grand thermodynamic potential
\begin{align}
\label{eq:Omega_nu}
&	\Omega_\nu(\rb)=- 2k_{B}T\int\frac{d^3\kb}{(2\pi)^3} \log\left[ 1+ 
e^{\bigl(\mu -e\phi(\rb) -E_\nu(\kb)\bigr)/k_{B}T} 
\right].
\end{align}
The subscript $\nu$ enumerates the hole energy bands, $\rb$ is the spatial coordinate, and $\phi(\rb)$ is the electrostatic potential; the factor 2 comes from the spin degeneracy. We consider two- and three-band models.

The hole energy $E_\nu(\kb)$ depends on strain and thus also on $\rb$. For the hole densities of interest and for not too low temperatures the major contribution to $\Omega_\nu$ comes from the wave vectors $\kb$ where strain-induced corrections to $E_\nu(\kb)$ are a perturbation. In this case we can expand $E_\nu$ in a series in the strain tensor $\hat\ep$,
\begin{align}
\label{eq:E_expansion_MD}
&E_\nu(\kb) = E_\nu\0 +
\sum_{n=1} E_\nu^{(n)}(\kb,\hat\ep), 
\end{align}
Here and below we use a superscript to show the order of the corresponding term in $\hat\ep$. Respectively, the functions $E_\nu^{(n)}$ in Eq.~(\ref{eq:E_expansion_MD}) scale as $\lVert\hat\ep\rVert^n$. They can be written in the form
\begin{align}
\label{eq:E_series}
E_\nu^{(n)}(\kb, \hat\ep) = \frac{1}{n!}\widehat U_\nu^{(n)}(\kb)\cdot\hat\ep^n.
\end{align}
The tensor $\widehat U_\nu^{(n)}$ here is independent of strain and is contracted with the tensor product of $n$ second-rank tensors $\hat\ep$. The powers $\hat\ep^n$ are defined recursively as 
\begin{align}
\label{eq:ep_powers}
 \hat \ep^n = \hat\ep\otimes \hat\ep^{n-1}, \quad \hat\ep^0 = \hat I.
\end{align}
The tensors $\widehat U_\nu^{(n)}$ can be calculated from the full Hamiltonian of the holes $\hat H(\kb,\hat\ep)$ using a direct perturbation theory. Below we will consider either the two-band approximation, in which one takes into account only light and heavy-hole bands, or the three-band approximation, which includes the split-off band. In the both cases the effect of the strain is taken into account by adding linear in $\hat\ep$ term to the unperturbed Hamiltonian, with proper symmetry. The full Hamiltonian constructed this way 
\[\hat H(\kb,\hat\ep)= \hat H\0(\kb) + \hat H_i\cdot\hat\ep\] 
is sometimes called the Luttinger-Kohn-Bir-Pikus Hamiltonian \cite{Terrazos2021}, see Appendix~\ref{sec:three-band}. The operators $\hat H\0$ and $\hat H_i$
are independent of coordinates. The matrix $\hat H\0$ is bilinear in the components of $\kb$, while $\hat H_i$ is independent of $\kb$, but both $\hat H\0$ and $\hat H_i$ have a similar structure, which is dictated by the symmetry. The tensors $\hat\ep$ are determined by the phononic displacement field. In the analysis of the direct effect of the coupling to holes on the eigenmodes the typical scale of the coordinate dependence of $\hat\ep$ is the size of the crystal.

Generally, the corrections to the energy $E_\nu\un$ can be found by a standard direct perturbation theory. We use this approach for the model where the split-off band is disregarded. For the $6\times 6$ Hamiltonian that takes this band into account, for numerical reasons, we use a different approach, see Section~\ref{subsec:method_3band}. 

To describe the mode frequency shift and the nonlinearity we need to find $E_\nu\un$ for $n=1,..,4$. The perturbation theory for $E_\nu\un$ diverges for $k\to 0$, where the splitting of the light- and heavy-hope bands goes to zero. However, the range of $k$ where the perturbation theory fails makes a negligible contribution to the parameters we are studying, and moreover, the diverging terms cancel each other in the expressions for the observables we are interested in. We show this analytically in Appendix~\ref{sec:divergence} for the model where the split-off band is disregarded, and we show this numerically where we take into account the split-off band.

\subsection{Free energy of the hole gas}
\label{subsec:free_energy}

Equations (\ref{eq:Omega_nu}) and (\ref{eq:E_expansion_MD}) allow one to calculate the hole free energy $F{} = \int d\rb \mathcal{F}{}(\rb)$ in the presence of strain, $\mathcal{F}{}(\rb) = \sum_\nu \Omega_\nu(\rb) + \bigl(\mu- e\phi(\rb)\bigr)\nh$, where $\nh$ is the local hole density. The strain-induced change $\Delta \mathcal{F}{}(\rb;\hat\ep)$ of the free energy density can be also expanded in a series in the powers of the strain tensor $\hat\ep$,
\begin{align}
\label{eq:delta_F_series}
\Delta \mathcal{F}{}(\rb;\hat\ep) = \sum_{n=1}
\Delta \mathcal{F}{}^{(n)}, \quad \Delta \mathcal{F}{}^{(n)} = \frac{1}{n!}\widehat\Lambda^{(n)}\cdot \hat\ep^n,
\end{align} 
where $\widehat\Lambda{}^{(n)}$ is a tensor of rank $2n$, and we again use central dot to indicate contraction with the $2n$-rank tensor $\hat\ep^n$. The tensors $\widehat\Lambda^{(n)}$ are determined by the tensors $\widehat U_\nu^{(m)}$ in the expansion of the band energy in $\hat\ep$ with $m\leq n$. They are independent of $\kb$ and of the coordinates in a spatially uniform crystal. They are symmetric, since $\hat\ep$ is a symmetric tensor and $\Delta \mathcal{F}{}(\rb)$ is an invariant. The number of independent components of the tensors $\widehat\Lambda^{(n)}$ in a cubic crystal for different ranks $2n$ is well-known, cf.~\cite{Landau1997,Nye2002}.

The hole-induced change of the mode eigenfrequencies is described by the tensor $\widehat\Lambda{}\2$, which gives the change of the elastic constants of the crystal. The leading-order terms in the dependence of the vibration frequency on the amplitude are determined by the quartic nonlinearity in $\hat\ep$, i.e., by the components $\widehat\Lambda{}\4$. There is also a contribution to this dependence of the terms $\widehat\Lambda{}\3$ in the second order of the perturbation theory, but as we show, it is comparatively small for the systems of interest. 

In the previous work on the effect of the hole-strain coupling \cite{Keyes1961,Bir1963,Cerdeira1972,Kim1976,Averkiev1984,Khan1985,Averkiev1987}, the analysis was done assuming electroneutrality and setting the potential $\phi(\rb)=0$. This is a good approximation for large values of the deformation potential parameters of the hole-strain coupling compared to $\mu$ and $k_BT$ and for a small Thomas-Fermi screening length. It allows for the effect of hole redistribution between the light- and heavy-hole bands and disregards the hole density change. We use this approximation in the present paper. 
The role of the strain-induced change of the acceptor and hole densities, the effect of the spatial nonunformity of doping, and other geometric effects will be analyzed in a separate paper. 

To calculate the tensors $\widehat \Lambda^{(n)}$ it is convenient to write the chemical potential as a series
\begin{align*}
\mu=\mu\0 + \sum_{n=1}\mu^{(n)}(\hat\ep), \quad \mu^{(n)} = \frac{1}{n!}\hat M^{(n)}\cdot \hat\ep{}^n
\end{align*}
and then expand $\Omega_\nu(\rb)$ in $\sum_{n=1}(\mu^{(n)} - E_\nu^{(n)})$. Here $\mu\0$ is the chemical potential in the absence of strain. The components $\mu\un $ are found from the condition 
\begin{align}
\label{eq:mu_condition}
-\frac{\partial}{\partial \mu}\sum_\nu\Omega_\nu = \nh,
\end{align}
where $\nh$ is the unperturbed hole density. This condition means that, in the expansion of $\partial_\mu\sum_\nu\Omega_\nu$, the terms $\propto \hat\ep{}^m$ with $m\geq 1$ should be equal to zero. In particular, we immediately obtain from the first-order perturbation theory
\begin{align}
\label{eq:first_order}
\mu\1 = a \tr\, \hat\ep,\quad \Delta\mathcal{F}\1 = \widehat\Lambda\1\cdot\hat\ep = a \nh \tr\,\hat\ep,
\end{align}
Here $a = D_H^{-1}\tr\, \hat H_i$; $D_H$ is the dimension of the Hamiltonian, $D_H=4$ and 6 for the two- and three-band models, respectively. The proportionality of $\mu\1$ and $\Delta\mathcal{F}\1$ to $\tr\,\hat\ep$ is just a consequence of the symmetry, as $\tr\,\hat\ep$ is the only linear in $\hat\ep$ scalar in a cubic crystal, see Appendix~\ref{sec:free_energy} for more details. The expressions for $\mu\un$ and $\widehat\Lambda{}\un$ with $n>1$ in terms of $E_\nu^{(m)}$ with $m\leq n$ are also given in Appendix~\ref{sec:free_energy}.

Equation~(\ref{eq:first_order}) for $\Delta\mathcal{F}\1$ describes the isotropic stress that results from the hole-strain coupling. The stress is compensated by the change of the volume of the crystal due to doping. This is similar to a linear in $\hat \ep$ term in the free energy of the host crystal, which is proportional to the change of temperature and describes thermal expansion \cite{Landau1986}. The volume changes also because of the very fact of doping, as the size of the substitutional atoms (acceptors) differs from the size of the host atoms. However, for a low dopant density, the change should be proportional to this density and thus be small. The nonlinear in $\hat\ep$ effect of the hole-strain coupling leads to a stronger effect. This is due to the large ratio of the coupling energy to the hole kinetic energy and, consequently, a large change of the statistical distribution of the holes by strain which, in turn, affects the strain. 

An advantageous feature of the representation (\ref{eq:E_expansion_MD})--(\ref{eq:delta_F_series}) for the analysis of the effect of the hole-strain coupling is that the free energy increment (\ref{eq:delta_F_series}) can be calculated directly for a given eigenmode $\ka$ as a function of its amplitude $A_\ka$. It is convenient to represent the displacement field of the mode $\ka$ as $A_\ka \ub_\ka(\rb)$. Here $\ub_\ka(\rb)$ is the dimensionless vector that describes the spatial profile of the mode $\ka$ and is normalized to the volume $V$ of the resonator, 
\[\int d^3\rb\, \ub_\ka^2 = V.\]
The time-dependent field is $A_\ka \ub_\ka(\rb) \cos\omega_\ka t$, where $\omega_\ka$ is the mode eigenfrequency. The strain $\hat\ep$ created by the mode has the form 

\begin{align}
\label{eq:kappa_strain}
\hat\ep = A_\ka\hat \ep_\ka \cos\omega_\ka t, \quad (\ep_\ka)_{ij} = \frac{1}{2}\left[\frac{\partial (u_\ka)_i}{\partial x_j} + \frac{\partial (u_\ka)_j}{\partial x_i}\right].
\end{align}
The subscripts $i,j$ run over the coordinate axes $x,y,z$; here and below we choose these axes along the $\braket{100}$ crystalline axes.

Replacing $\hat\ep$ with $A_\ka\hat\ep_\ka \cos\omega_\ka t $ determines the ``instantaneous'' change of the hole energy $E_\nu\un(\kb,\hat\ep)$ and the free energy $\Delta\mathcal{F}(\rb;\hat\ep)$ due to the mode $\ka$. Disregarding delay is justified for low-frequency eigenmodes, since the associated strain is adiabatically followed by the holes, as explained above. In turn, the change of $\Delta\mathcal{F}$ calculated this way determines the change of the eigenfrequency of the mode $\ka$ and the mode nonlinearity due to the hole-strain coupling.


\section{Mode eigenfrequencies in the two-band approximation}
\label{sec:T_dependence}

The effect of the hole-strain coupling on the mode eigenfrequencies comes from the coupling-induced change of the linear elasticity parameters \cite{Landau1986}, i.e., the terms $\propto \hat\ep^2$ in the free energy. Thus the change is determined by $\Delta\mathcal{F}\2$ in Eq.~(\ref{eq:delta_F_series}) and is described by the tensors $\widehat\Lambda\2$. As seen from Eqs.~(\ref{eq:Omega_nu})--(\ref{eq:delta_F_series}), it depends on the hole density and temperature. 
For the typical hole densities, the doping-induced change of the eigenfrequencies is small. However, it can be comparable to the ``intrinsic'' temperature-dependent eigenfrequency change due to the lattice nonlinearity, i.e., the change in the absence of doping. The overall temperature dependence of the mode eigenfrequency is a sum of the terms that come from the doping and the lattice nonlinearity.


\subsection{Estimate of the mode frequency change}
\label{subsec:estimate}

It is useful to have an estimate of the effect. The value of $\widehat\Lambda\2 \cdot \hat\ep^2 $ is determined by the strain-induced change of the hole energies $E_\nu\2$ multiplied by the hole density $\nh$. The characteristic value of $E_\nu\2$ can be estimated from the second-order perturbation theory in the hole-strain coupling. If the characteristic coupling constant (the characteristic parameter of the deformation potential) is $\mathcal{D}$, the conventional estimate is
\[E_\nu\2 \sim (\mathcal{D}^2/E_\mathrm{kin}) \lVert \hat\ep\rVert^2,\]
where $E_\mathrm{kin}$ is the typical hole kinetic energy, which varies from $\mu_0$ for the strongly degenerate hole gas to $k_BT$ for the nondegenerate gas. 

For the strain given by Eq.~(\ref{eq:kappa_strain}) we have $\lVert \hat\ep\rVert^2 \sim A_\ka^2/L^2$ where $L$ is the typical size on which the mode displacement ${\bf u}_\ka$ varies; this can be the length of the crystal, for an extensional mode, or the size of the square-shape crystal, for a Lam\'e mode. Therefore, to the second order in $\hat\ep$, the overall change of the free energy $F_\ka\2$ for the mode $\ka$ is 
\[F_\ka\2 \sim (\mathcal{D}^2/E_\mathrm{kin})V \nh A_\ka^2/L^2,\]
where $V$ is the volume of the crystal. 

This change has to be compared with the typical energy of the mode in the absence of the coupling,
\[M\omega_\ka^2 A_\ka^2 \sim \rho V A_\ka^2 v_s^2/L^2,\]
where $M$ is the mass of the crystal, $\rho$ is the crystal density, and $v_s$ is the sound velocity. 

The ratio 
\begin{align}
\label{eq:estimate_ratio}
\zeta =F_\ka\2/(M\omega_\ka^2 A_\ka^2) \sim (\mathcal{D}^2/E_\mathrm{kin})\nh /\rho v_s^2
\end{align}
determines the coupling strength. For typical $\mathcal{D} \sim 3$~eV, $\nh\sim 10^{19}\,\mathrm{cm}^{-3}$, $E_\mathrm{kin}\sim 0.03$~eV, $\rho \sim 3\,\mathrm{g/cm}^3$, and $v_s\sim 5\times 10^5$~cm/s we obtain $\zeta\sim 0.006$. The value of $\zeta$ also provides an estimate of the change of the mode frequency due to the coupling to holes, $\Delta\omega_\ka \sim \zeta\omega_\ka$. Even though the frequency change is small, it can have a profound temperature dependence and thus significantly affect the overall temperature dependence of the frequency. This is important for various applications of mechanical resonators, cf. \cite{Beek2011,Ng2013}.

By construction, the $n$th order tensor $\widehat\Lambda^{(n)}$ in the expansion of the free energy (\ref{eq:delta_F_series}) scales as $ \mathcal{D} \bigl(\mathcal{D}/E_\mathrm{kin}\bigr)^{n-1}$. 
Therefore $\lVert\widehat\Lambda^{(n)}\rVert$ increases fast with the increasing order $n$. This is why the hole-strain coupling makes a major contribution to the mode nonlinearity. For the same reason, in $\Delta\mathcal{F}$, the leading-order contribution to the $n$th order term in the displacement $\ub$ comes from 
the term $[(\partial u_i/\partial x_j) + (\partial u_j/\partial x_i)]/2$ in $\ep_{ij}$ whereas the nonlinear term $(\partial u_k/\partial x_i)(\partial u_k/\partial x_j) $ in $\ep_{ij}$ can be disregarded, as it was done in Eq.~(\ref{eq:kappa_strain}).


\subsection{Two-band energy spectrum}
\label{sec:dispersion_law}

For not too large hole densities and not too high temperatures, the split-off band is empty, and it is sufficient to take into account the bands of light and heavy holes only. The subscript $\nu$ in Eqs.~(\ref{eq:Omega_nu}) and (\ref{eq:E_expansion_MD}) takes on values 0 and 1, which we relate to the light-and heavy-hole bands, respectively. The hole dispersion law with the account taken of the hole-strain coupling is known in the explicit form for this case \cite{Bir1974,Seeger2004}. This allows obtaining expressions for the corrections to the frequencies of eigenmodes in the form of simple integrals. 

The energy dispersion law in the absence of hole-strain coupling is $E_\nu\0(\kb)=Ak^2 + (-1)^\nu \mathcal{E}(\kb)$, where
\begin{align}
\label{eq:E_lh_hh_zero}
&	\mathcal{E}(\kb)=\left(
B^2k^{4}+\frac{1}{2}C^2\sum_{i\neq j}k_i^2k_j^2\right)^{1/2}.
\end{align}
Parameter $C$ in this expression characterizes the warping of the energy surfaces. 

In the deformation potential approximation the dispersion law becomes 
\begin{align}
\label{eq:strained_dispersion}
	&E_\nu(\kb)=Ak^2+a\tr\hat{\ep}+(-1)^\nu \left(\mathcal{E}^2(\kb)+\Xi\1+\Xi\2\right)^{1/2},
\nonumber\\
&\Xi\1=Bb\left(3\sum_i k_i^2\ep_{ii}-k^2\tr\hat{\ep}\right)
+Dd\sum_{i\neq j} k_ik_j \ep_{ij},
\nonumber\\
&\Xi\2=\sum_{i\neq j} \left[ \frac{b^2}{4} (\ep_{ii}-\ep_{jj})^2 + \frac{d^2}{2}\ep_{ij}^2\right],
\end{align}
where the superscripts $1$ and $2$ refer to the terms that are linear and quadratic in $\hat\ep$, respectively.

The parameters $a,b,$ and $d$ in Eq.~(\ref{eq:strained_dispersion}) are the deformation potential parameters. They determine the strength of the hole-strain coupling and, typically, are in the range of a few electron volts, much larger than the chemical potential and $k_BT$. An important feature of the two-band approximation is that, as seen from Eq.~(\ref{eq:strained_dispersion}), the band splitting by strain is ``symmetric'': the light and heavy-hole bands shift in the opposite direction by the same amount, for a given strain.


\subsection{Analytical expressions}
\label{subsec:analytical_warped}

Equation~(\ref{eq:strained_dispersion}) for $E_\nu$ allows one to find the components $E_\nu\un$ in the series (\ref{eq:E_series}) simply by expanding $E_\nu$ in $\Xi\1$ and $\Xi\2$. The first- and second-order terms in $\hat\ep$ are
\begin{align}
\label{eq:E_lh_hh_second}
&
E_\nu\1(\kb,\hat\ep) = a\tr\,\hat\ep +(-1)^\nu \Xi\1/2\mathcal{E}(\kb),\nonumber\\	
&	
E_\nu\2(\kb,\hat\ep)= (-1)^\nu \left(\frac{\Xi\2}{2\mathcal{E}(\kb)}-\frac{\Xi^{(1) 2}}{8\mathcal{E}(\kb)^3}\right).
\end{align}

To find $\widehat\Lambda\2$, one has to find the terms up to the second order in $\hat\ep$ in $\Omega_\nu$ and $\mu$. The expressions that relate $\Omega_\nu$ and $\mu$ to $E_\nu\un$ are given in see Appendix~\ref{sec:free_energy}. The calculation for the two-band model is facilitated by the relation
\begin{align}
\label{eq:E1_symmetry_two_band}
& \int d^3\kb \left[\frac{\Xi\1(\kb,\hat\ep)}{\mathcal{E}(\kb)}\right] f(E_\nu\0,\mu\0)=0,\nonumber\\
& f(E_\nu,\mu) = \left\{\exp\left[(E_\nu-\mu)/k_BT\right]+1\right\}^{-1}
\end{align}
that follows from the symmetry arguments. This relation shows, in particular, that $\Omega_\nu\1=0$ and leads to Eqs. (\ref{eq:first_order}) for $\mu\1$ and $\Delta\mathcal{F}\1$; Eqs. (\ref{eq:first_order}) imply that $
\widehat\Lambda{}^{(1)}_{ij} = a \nh\delta_{ij}$.
%

A straightforward calculation of $\widehat\Lambda\2$ based on the results of Appendix~\ref{sec:free_energy} gives

\begin{widetext}
\begin{align}
\label{eq:Lambda_11_MD}
\Lambda{}_{1111}\2=-2\Lambda{}_{1122}\2 =
	4 \sum_\nu\int\frac{d^3\kb}{(2\pi)^3}
	\left\{(-1)^\nu f_\nu
	\left[ \frac{b^2}{2\mathcal{E}(\kb){}} - \frac{B^2b^2(9k_x^4 - k^4)}{8\mathcal{E}(\kb)^3}\right]
	- f'_\nu\frac{B^2b^2(9k_x^4 - k^{4})}{8\mathcal{E}(\kb)^2}
	\right\}
\end{align}
and
\begin{align}
\label{eq:Lambda_12_MD}
	\Lambda{}_{1212}\2= \sum_\nu \int\frac{d^3\kb}{(2\pi)^3}
	\left\{(-1)^\nu f_\nu
	\left[\frac{d^2}{2\mathcal{E}(\kb)} - \frac{D^2d^2k_{x}^2k_{y}^2}{2\mathcal{E}(\kb)^3}\right]
	- f'_\nu \frac{D^2d^2k_{x}^2k_{y}^2}{2\mathcal{E}(\kb)^2}
	\right\}.
\end{align}
\end{widetext}
Here and in the following, we use the notation
\[f_\nu \equiv f(E_\nu\0,\mu\0),\quad f'_\nu \equiv \partial f_\nu/\partial \mu\0\]
and use an appropriate number of primes to denote higher-order derivatives of $f_\nu$ with respect to $\mu\0$ (cf. Appendix~\ref{sec:free_energy}).
In often used Voigt notation, $\Lambda{}_{1111}\2 = c_{11}^{(\mathrm{h})}$, $\Lambda{}_{1122}\2 = c_{12}^{(\mathrm{h})}$, and $\Lambda{}_{1212}\2 = 2c_{44}^{(\mathrm{h})}$, where $c_{ij}^{(\mathrm{h})}$ are the doping-induced terms in the corresponding elasticity parameters. We note that, generally, there are three independent components of the linear elasticity tensor in a cubic crystal, but the hole-induced renormalization of this tensor $\widehat\Lambda\2$ has only two parameters.

Equations (\ref{eq:Lambda_11_MD}) and (\ref{eq:Lambda_12_MD}) provide an analytical solution of the problem of free energy of holes in the presence of strain with the account taken of the warping of the energy surfaces and of the associated change of the elasticity parameters. In contrast to the previous analysis \cite{Bir1963}, where the warping was disregarded, the integrals over $\kb$ cannot be evaluated in the explicit form. Since $\mathcal{E}(\kb)\propto k^2$, the integrals converge at small $k$, whereas at large $k$ the convergence is guaranteed by the exponential falloff of $f_\nu$. 
In the limit where $\mu\0\gg k_BT$ the parameters $\hat\Lambda\2$ scale as $n_\mathrm{h}^{1/3}$. The dependence of $\hat\Lambda\2$ on the hole density and on the temperature becomes fairly complicated in the most interesting range of doping and temperatures where $\mu\0/k_BT \sim 1$. This is due to the complicated form of the $\kb$-dependent factors in Eqs.~(\ref{eq:Lambda_11_MD}) and (\ref{eq:Lambda_12_MD}). Therefore further analysis requires numerical calculations. However, the calculations are straightforward given the explicit form of the expressions for the doping-induced change of the elasticity parameters.


\subsection{Numerical results}
\label{subsec:T_dependence}

\begin{figure*}
	\centerline{\includegraphics[width=0.9\textwidth]{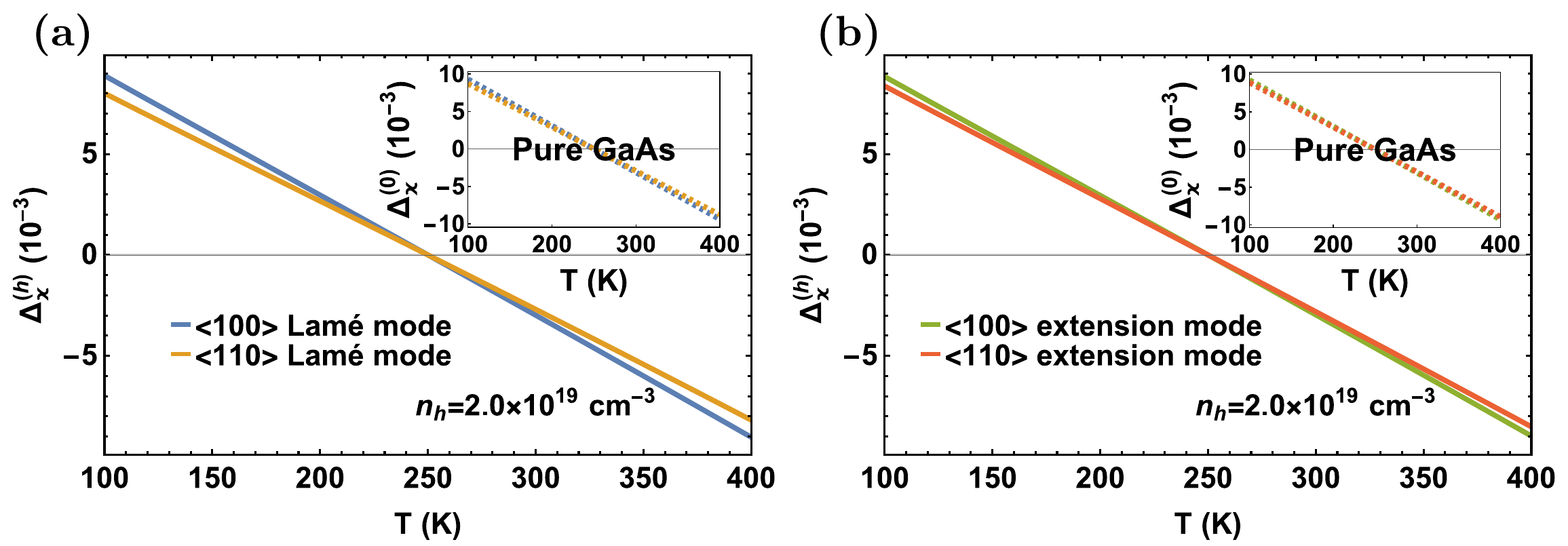}}
	\protect\caption{Temperature dependence of the relative linear frequency shifts $\Delta_\ka^{(\mathrm{h})}=[\omega_{\ka}(T)-\omega_{\ka}(T_{0})]/\omega_{\ka}(T_{0})$ at $T_{0}=250$ K for (a) Lam\'e modes and (b) extension modes in GaAs-based NEMS or MEMS at $\nh=2.0\times10^{19}$ cm$^{-3}$. The insets show the results for undoped GaAs. The data on the temperature dependence of the intrinsic second-order elastic constants in GaAs is taken from Ref. \cite{jordan1980evaluation}.}
	\label{fig:GaAs_two_band_linear}
\end{figure*}

In this section, we use Eqs.~(\ref{eq:Lambda_11_MD}) and (\ref{eq:Lambda_12_MD}) to describe the coupling-induced dependence of the eigenfrequencies of NEMS and MEMS modes on temperature and hole density. The numerical integration over $\kb$ is done using the standard Mathematica package. We provide results for the Lam\'e and extensional modes, as such modes are frequently studied in the experiments \cite{Ng2015,Rodriguez2019,Lee2016}.

For the Lam\'e modes in square plates cut out along $\braket{100}$ and $\braket{110}$ axes the dependence of the eigenfrequencies on the elastic constants (in Voigt notation) is, respectively,
\begin{align}
\label{eq:mode_elasticity}
	&\omega_{L,\braket{100}} \propto C_{L,\braket{100}}(T)=\sqrt{c_{11} - c_{12}}, \nonumber\\
	&\omega_{L,\braket{110}}\propto C_{L,\braket{110}}(T)=\sqrt{c_{44}}.
\end{align}
For a long narrow beam cut along $\braket{100}$ axis with the sides parallel to $(100)$ planes the eigenfrequency of the extension mode scales as
\begin{align*}
	&\omega_{e,\braket{100}} \propto C_{e,\braket{100}}(T)\\
	&=\sqrt{[c_{11}(c_{11}+ c_{12})-2c_{12}^2]/(c_{11} + c_{12})},
\end{align*}
whereas for an extensional mode along $\braket{110}$ axis, with the beam sides parallel to the $(001)$ and $(1\bar{1}0)$ planes 
\begin{align*}
	&\omega_{e,\braket{110}}\propto C_{e,\braket{110}}(T)\\
	&=\sqrt{\frac{c_{44}\left[c_{11}\left(c_{11}+c_{12}\right)-2c_{12}^2\right]}{c_{11}\left(c_{11}+c_{12}+2c_{44}\right)-2c_{12}^2}},
\end{align*}
cf.~\cite{graff1991wave}. 
The combinations $C_{e,\braket{100}}$ and $C_{e,\braket{110}}$ also determine the temperature dependence of the hole-induced corrections to the frequencies of flexural modes of the appropriately oriented beams, since these frequencies depend on the flexural rigidity which, in turn, depends on the Young modulus \cite{Landau1986}.

The two-band model considered in this section applies in an appreciable range of densities and temperatures to resonators made out of Ge and III-V semiconductors, in which the spin-orbit splitting of the valence band is comparatively large. 
In contrast, in Silicon the spin-orbit splitting is small and the applicability of the two-band model is limited to very small hole densities and low temperatures.

Here, we present results for GaAs-based micromechanical resonators. Such resonators are widely used for sensing forces of various natures, from biomolecular to magnetic, as well as in accelerometers and other devices. They are advantageous in several respects \cite{Yamaguchi2017}. In particular, because GaAs is piezoelectric, mechanical vibrations can be directly controlled and measured through strain-voltage transduction. Besides applications, GaAs-based NEMS and MEMS are also broadly used in fundamental studies, cf. \cite{Yamaguchi2017, Hamoumi2018,*Allain2021} and references therein.

In Fig. \ref{fig:GaAs_two_band_linear} we show the temperature dependence of the relative linear frequency shifts of different eigenmodes of GaAs resonators in the two-band approximation for the hole density $n_\mathrm{h}=2.0\times10^{19}$ cm$^{-3}$. Plotted is the relative increment $\Delta_\ka^\mathrm{(h)}$ of the eigenfrequency $\omega_\ka(T)$ counted off from the value $\omega_\ka(T_0)$ at a certain reference temperature $T_0$: 
\begin{align}
\label{eq:relative_shift}
\Delta_\ka^\mathrm{(h)}=\frac{\omega_{\ka}(T)-\omega_{\ka}(T_{0})}{\omega_{\ka}(T_{0})}
=\frac{C_{\ka}(T)-C_{\ka}(T_{0})}{C_{\ka}(T_{0})}.
\end{align}
The temperature effects of doping and the intrinsic nonlinearity are small and could be added together. We used the values of the elasticity parameters $c_{ij}$ in the absence of doping provided in \cite{varshni1970temperature} and the valence band parameters given in \cite{winkler2003spin}.
In calculating this increment we took into account both the doping-induced change of the elasticity parameters described by Eqs.~(\ref{eq:Lambda_11_MD}) and (\ref{eq:Lambda_12_MD}) and the temperature dependence of these parameters due to the intrinsic nonlinearity of the crystal. The effect of the intrinsic nonlinearity 
was measured for different levels of $n$-type doping \cite{burenkov1973temprature,cottam1973elastic} and was found to weakly depend on such doping \cite{jordan1980evaluation}.

Figure~\ref{fig:GaAs_two_band_linear} shows that $p$-type doping does affect the temperature dependence of the elastic constants, but the effect is comparatively small. The dependence of the mode eigenfrequencies on $T$ remains monotonic and is close to linear even for a comparatively large doping. We did calculations for several other densities $n_\mathrm{h}$, with $n_\mathrm{h}$ in the range of $10^{18}$--$10^{20}~\mathrm{cm}^{-3}$; the doping-induced change of the parameters is linear in $n_\mathrm{h}$ in this range and the dependence on $T$ remains close to linear. As we will see, the results for Si are qualitatively different.


\section{Nonlinearity in the two-band approximation}
\label{sec:nonlinearity}

An important characteristic of the eigenmodes in nano- and micromechanical resonators is their nonlinearity. Because of the nonlinearity, the vibration frequency depends on the vibration amplitude $A_\ka$ \cite{Landau2004a}. This has several profound consequences. One of them is that fluctuations of the amplitude, for example those due to thermal noise, lead to frequency fluctuations and broadening of the power spectrum, see \cite{Dykman1984}. Such broadening has been seen in various types of nano- and micromechanical systems \cite{Barnard2012,Venstra2012,Matheny2013,Gieseler2013,Miao2014,Maillet2017,Huang2019,Amarouchene2019,Huang2019}, and see \cite{Bachtold2022a} for a review. Another important consequence is that the nonlinear frequency change limits the amplitude range used in the devices that rely on frequency stability.

The hole-strain coupling can significantly modify the mode nonlinearity. This was indicated already in Refs.~\cite{Keyes1961,*Keyes1968}. The effect is a consequence of the large ratio of the deformation potential parameter $\mathcal{D}$ to the typical hole kinetic energy $E_\mathrm{kin}$; for the two-band model considered here $\mathcal{D}=\mathrm{max}(b,d)$. We recall that, in the expression for the free energy (\ref{eq:delta_F_series}), $\widehat\Lambda^{(n)}\propto \mathcal{D} \bigl(\mathcal{D}/E_\mathrm{kin}\bigr)^{n-1}$. In many semiconductors $\mathcal{D}/E_\mathrm{kin} \gtrsim10^2$ for the hole densities $10^{17}$--$10^{19}$~cm$^{-3}$ and room temperatures. Therefore $\lVert\widehat\Lambda^{(n)}\rVert$ is quickly increasing with the increasing $n$, as indicated earlier.

For not too large amplitudes $A_\ka$, the hole contribution to the dependence of the frequency $\omega_\ka$ of the mode $\ka$ on its amplitude $A_\ka$ has the form \cite{Dykman1984}
\[\delta\omega_\ka= 
\frac{3\gamma_\ka^\mathrm{(h)}}{8 M\omega_\ka} A_\ka^2.\]
Here $\gamma_\ka^\mathrm{(h)}$ is the hole contribution to the parameter of quartic (Duffing, or equivalently, Kerr) nonlinearity of the potential energy of the mode. In terms of the normal coordinate $Q_\ka$ of the mode, this potential energy, with the account taken of the quartic nonlinearity, is $\frac{1}{2} M\omega_\ka^2 Q_\ka^2 + \frac{1}{4}\gamma_\ka^\mathrm{(h)} Q_\ka^4$). The parameter $\gamma_\ka^\mathrm{(h)}$ is determined by the quartic term $\Delta\mathcal{F}\4$ in the free energy with respect to the normalized lattice displacement field $\ub_\ka(\rb)$ that describes the spatial profile of the mode $\ka$, 
\begin{align}
\label{eq:gamma_lambda}
&\gamma_\ka^\mathrm{(h)} = 4\int d\rb \,\Delta \mathcal{F}{}\4_\ka(\rb)\nonumber\\
&\equiv\frac{1}{6}\int d\rb \widehat\Lambda\4 \cdot \hat\ep_\ka\otimes\hat\ep_\ka\otimes\hat\ep_\ka\otimes\hat\ep_\ka,
\end{align}
where the strain tensor $\hat \ep_\ka$ for mode $\ka$ is defined in Eq.~(\ref{eq:kappa_strain}).

Generally speaking, $\gamma_\ka^\mathrm{(h)}$ has a contribution from the square of the terms $\propto \widehat\Lambda\3$. This contribution can be separated into two parts. One part comes from the ``self-action'' of the mode mediated by the coupling to holes. Such self-action gives a term $\frac{1}{3}\beta_\ka Q_\ka^3$ in the mode potential energy, with $\beta_\ka\propto \lVert \hat\Lambda\3\rVert$. The contribution of this term to $\gamma_\ka^\mathrm{(h)}$ is $\propto \beta_\ka^2/M\omega_\ka^2$ \cite{Landau2004a} and thus is $\sim \zeta \gamma_\ka^\mathrm{(h)}$, where $\zeta$ is given by Eq.~(\ref{eq:estimate_ratio}). Since $\zeta\ll 1$ in the systems we consider, it can be disregarded. The second part comes from the hole-induced nonlinear coupling of the considered mode to other modes. It has the same smallness; however, it gives an important contribution to the decay rate of the mode, as discussed in Section~\ref{sec:Akhiezer}.

The value of $\Delta \mathcal{F}{}\4_\ka$ is obtained by expanding $\Omega_\nu$ and $\mu$ to the fourth and the third order in $\hat\ep$, respectively. Such expansion is given in Appendix~\ref{sec:free_energy}. The general expressions can be further simplified because of the structure of $E_\nu\1$ in the two-band approximation, see Eq.~(\ref{eq:E_lh_hh_second}). In particular, the term $\propto \mu\3$ drops out from the expression for $\Delta \mathcal{F}{}\4_\ka$. Therefore it is sufficient to find only $\mu\2$. 

It is seen from the explicit expression for the hole energy in the presence of strain $E_\nu(\kb)$, Eq.~(\ref{eq:strained_dispersion}), that the expansion in $\hat\ep$ is an expansion in $\hat\ep/k^2$. For small wave numbers $k$, the individual terms in the integrals over $\kb$, which give $\Delta\mathcal{F}\4$, diverge as $k^{-6}, k^{-4}$, and $k^{-2}$. However, as shown in Appendix~\ref{sec:divergence}, the diverging terms cancel each other, so that the overall expression for $\Delta\mathcal{F}_\ka\4$ is free from divergences.

We illustrate the effect of the coupling-induced nonlinearity for the Lam\'e modes in GaAs square plates with side $L$ and thickness $h$. The doping-related nonlinearity parameter $\gamma_\ka^\mathrm{(h)}$ is related in a simple way to the fourth-order elasticity parameters \cite{moskovtsev2017strong}. 
For the plates cut out along $\braket{100}$ and $\braket{110}$ crystal symmetry axes
\begin{align*}
	\gamma_{L,\braket{100}}=\left(\frac{27\pi^4 h}{32L^2}\right)c_{1111}^{(\text{h})}
\end{align*}
and
\begin{align*}
	\gamma_{L,\braket{110}}=\left(\frac{3\pi^4 h}{2L^2}\right)c_{4444}^{(\text{h})}.
\end{align*}
Here, $c_{1111}^{(\text{h})}$ and $c_{4444}^{(\text{h})}$ are the doping-related corrections to the nonlinear elasticity coefficients in the Voigt notation.

\begin{figure}[htp!]
	\centerline{\includegraphics[clip, width=0.9 \columnwidth]{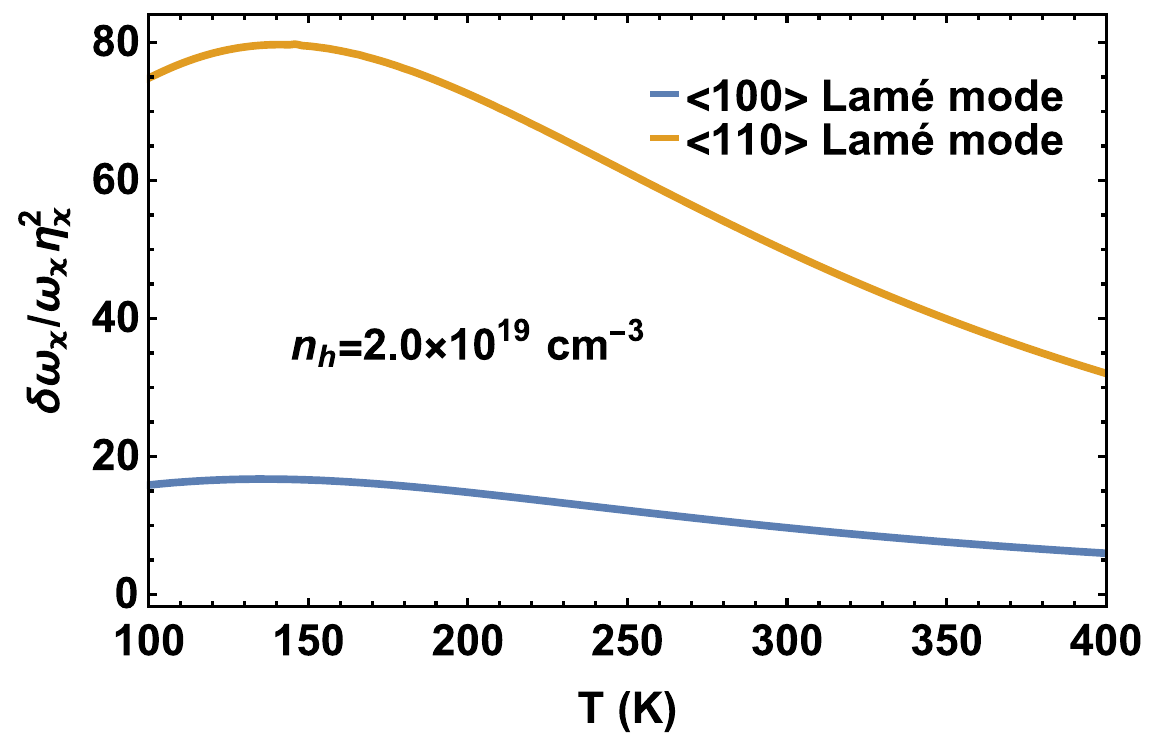}}
	
	\protect\caption{Temperature dependence of the scaled relative nonlinear frequency shifts $\delta \omega_\ka/\omega_\ka \eta_{\ka}^{2}$ for $\braket{100}$ and $\braket{110}$ Lam\'e modes in a square single-crystal GaAs resonator for the hole density $\nh=2.0\times10^{19}$~cm$^{-3}$.}
	
	\label{fig:GaAs_two_band_nonlinear}
\end{figure}

It is convenient to scale the vibration amplitude $A_{\ka}$ by the resonator size, $A_{\ka}=\eta_{\ka}L$.
The dimensionless parameter 
\begin{align}
\label{eq:scaled_nonlinear_param}
\aleph_\ka=\frac{\delta \omega_\ka}{\omega_\ka \eta_\ka^{2}} = \frac{3\gamma_\ka^\mathrm{(h)}L^2}{8 M\omega_\ka^2} , 
\end{align}
which depends only on the second- and fourth-order elasticity coefficients, characterizes the relative nonlinear frequency shift \cite{moskovtsev2017strong}. In Fig.~\ref{fig:GaAs_two_band_nonlinear} we show the temperature dependence of this parameter for the Lam\'e modes in two differently oriented resonators. The temperature dependence of the doping-related terms $c_{1111}^{(\text{h})}$ and $c_{4444}^{(\text{h})}$ is calculated in the two-band approximation using Eq.~(\ref{eq:F_fourth}). The results refer to the hole density $\nh=2.0\times10^{19}$~cm$^{-3}$. For this density, the ratio $\mu_0/k_BT$ varies from $\approx 4.4$ at 100~K to $\approx 1.3$ at 250~K and $\approx 0.27$ at 400~K.

The nonlinear frequency shift shown in Fig.~\ref{fig:GaAs_two_band_nonlinear} is a nonmonotonic function of temperature. This is a consequence of the structure of Eq.~(\ref{eq:F_fourth}) for the correction to the free energy $\Delta\mathcal{F}\4$ that gives $\gamma_\ka\uh$. This correction is a sum of the contributions from the integrals over $\kb$ of corrections to the hole energy $E_\nu^{(n)}(\kb)$, up to quartic order in $\hat\ep$ (i.e., $n=1,\dots,4$), weighted by the Fermi distribution $f_\nu$ and its first, second, and third derivatives. For $\mu_0/k_BT \gg 1$, where the hole gas is strongly degenerate, the leading-order contributions come from the terms that contain $f_\nu$ and $f_\nu'$. An important feature of the two-band model is that the hole-energy corrections $E_0^{(n)}(\kb)$ and $E_1^{(n)}(\kb)$ have opposite signs, and the linear in $E_\nu\un$ terms (these are terms $\propto E_\nu\4$, which are weighed by $f_\nu$) partly compensate each other. Therefore, $\aleph_\ka\propto \gamma_\ka\uh$ is comparatively small. With the increasing $T$, there is an increasing contribution to $\Delta\mathcal{F}\4$ from the terms quadratic in $E_\nu^{(n)}$, which are weighted by the derivatives of $f_\nu$ with respect to $\mu_0$. As a result, $\gamma_\ka\uh$ increases. On the other hand, at high temperatures, $\gamma_\ka\uh$ should decrease with the increasing $T$. This is because $E_\nu\un(\kb)$ is $\propto (\mathcal{D}\lVert\hat\ep\rVert/k_BT)^n$ for holes with thermal wave vectors $k_{\mathrm{th}}$ when $k_BT \gg |\mu_0|$.


\section{The effect of the split-off band}
\label{sec:spin_split_off_band}

The two-band model may be insufficient to describe the effects of doping at large hole densities and high temperatures. A better approximation is provided by the three-band model, which takes into account the split-off band along with the bands of light and heavy holes \cite{Luttinger1955}. This is particularly important for Silicon, where the spin-orbit coupling is relatively weak. The three-band model in the presence of strain is well described by the broadly used Luttinger-Kohn-Bir-Pikus Hamiltonian \cite{Bir1974}.

For completeness, we provide the three-band Hamiltonian $\hat H(\kb)$ in Appendix \ref{sec:three-band}. Its form is dictated by the symmetry of the crystal. It is a $6\times 6$ matrix, a sum of the Hamiltonian in the absence of strain $\hat H\0(\kb)$ and the coupling Hamiltonian $\hat H_i\cdot\hat\ep$, which is linear in strain. The eigenvalues of the Hamiltonian $E_\nu(\kb,\hat\ep)$ are double-degenerate due to spin. There are 3 eigenvalues for each $\kb$. The energy branches are enumerated by $\nu=0,1,2$, corresponding to the light-hole, heavy-hole, and split-off bands.


\subsection{The numerical method}
\label{subsec:method_3band}

To describe the effect of the hole-strain coupling in the three-band model we use the general formalism developed for the two-band approximation in Sections~\ref{sec:T_dependence} and \ref{sec:nonlinearity}. Implementing this formalism requires finding the energies $E_{\nu}(\kb,\hat\ep)$ and expanding them in powers of $\hat\ep$. The expansion coefficients $E_{\nu}\un(\kb)$ can be then used to calculate the series expansion of the free energy in $\hat \ep$; the corresponding expressions are given in Appendix \ref{sec:free_energy}. However, finding the eigenvalues $E_\nu(\kb,\hat\ep)$ of the $6\times 6$ Hamiltonian is analytically intractable. Therefore, we numerically evaluated $E_{\nu}(\kb,\hat\ep)$ and then found its expansion in $\hat\ep$. The integration over $\kb$ in Eqs. (\ref{eq:mu_second})--(\ref{eq:F_fourth}) was replaced by summation over a three-dimensional uniformly meshed grid in the $\kb$ space. The exponential falloff of the integrands at large $k$ allowed us to limit the integration domain, and the independence of the large-$k$ cutoff was tested. 

A conventional way of calculating the expansion coefficients $E_\nu\un(\kb)$ is based on the perturbation theory \cite{Landau1997}. However, it requires finding not only the eigenvalues, but also the eigenvectors of the operator $\hat H\0(\kb)$ for each $\kb$. Calculating the eigenvectors and then the matrix elements to the fourth order of the perturbation theory in $\hat H_i\cdot\hat\ep$ is numerically demanding.

Since we are interested in the energy changes for specific eigenmodes, we could use a different approach. For each $\kb$ on the grid we calculated $E_{\nu}(\kb,\hat\ep)$ for $\hat{\ep}=0$ and then for discretized values of the strain tensor for a given mode $\ka$, i.e., for $\pm \hat{\ep}_{\ka}$, $\pm2\hat{\ep}_{\ka}$, $\dots$. Then we used the standard finite-difference formulae \cite{fornberg1988generation} or $n$th-degree-polynomial interpolation to extract the expansion coefficients in the power series $\sum_{n}\widehat U_\nu^{(n)}(\kb)\cdot\hat{\ep}_{\ka}^{n}/n!$ for $E_\nu(\kb,\hat\ep_\ka)$, see Eq.~(\ref{eq:E_series}). We tested that both approaches gave the same result, to high accuracy, and the result was independent of the discretization step. The values of $E_{\nu}\un(\kb)$ found in this way for the strain $\hat{\ep}_{\ka}$ at each $\kb$ were then used to find the free energy for the corresponding strain. The results were further utilized to evaluate the effect of the coupling to holes on the corresponding eigenmodes of a resonator.

The coupling-induced corrections to the elasticity parameters that determine the eigenfrequencies of the Lam\'e modes were presented earlier in Ref.~\cite{Khan1985}. Our results agree with the results on the parameter $c_{11}\uh-c_{12}\uh$, which determines $\omega_{L,\braket{100}}$. For $c_{44}\uh$, which determines $\omega_{L,\braket{110}}$, the difference is close to $25\%$. Most likely, it comes from a more recent value \cite{winkler2003spin} of the deformation potential parameter $D_u$ that we are using; this parameter strongly affects $c_{44}\uh$. In contrast to Ref.~\cite{Khan1985}, we limited our analysis to not too low densities $\nh$ and not too low temperatures, where all acceptors are ionized. This is required to make the hole density independent of temperature, so that it is a well-controlled parameter. Indeed, the Thomas-Fermi screening length $[(4\pi e^2/\epsilon_\mathrm{Si})\partial \nh/\partial \mu_0]^{-1/2}$ is $\lesssim 1$ nm in Si at $\nh=2.0\times10^{19}$ cm$^{-3}$ and $T=300$ K. Given that the Bohr radius is $\gtrsim 2$ nm and the acceptor ionization energy is about $0.05$ eV \cite{kittel1954theory,kittel2005introduction}, it is a good approximation to assume that all acceptors are ionized for such $\nh$ and $T$. However, this assumption does not work in the range $\nh =10^{17}~\mathrm{cm}^{-3}$ and $T=100$~K, which was included in the analysis of Ref.~\cite{Khan1985}. Other important distinctions from \cite{Khan1985} are that, besides the Lam\'e modes, we consider extensional modes and, in the first place, study the mode nonlinearity.


\subsection{Mode eigenfrequencies}
\label{subsec:eigenfrequencies_3band}

\begin{figure}[htp!]
	\centerline{\includegraphics[width=1.0\columnwidth]{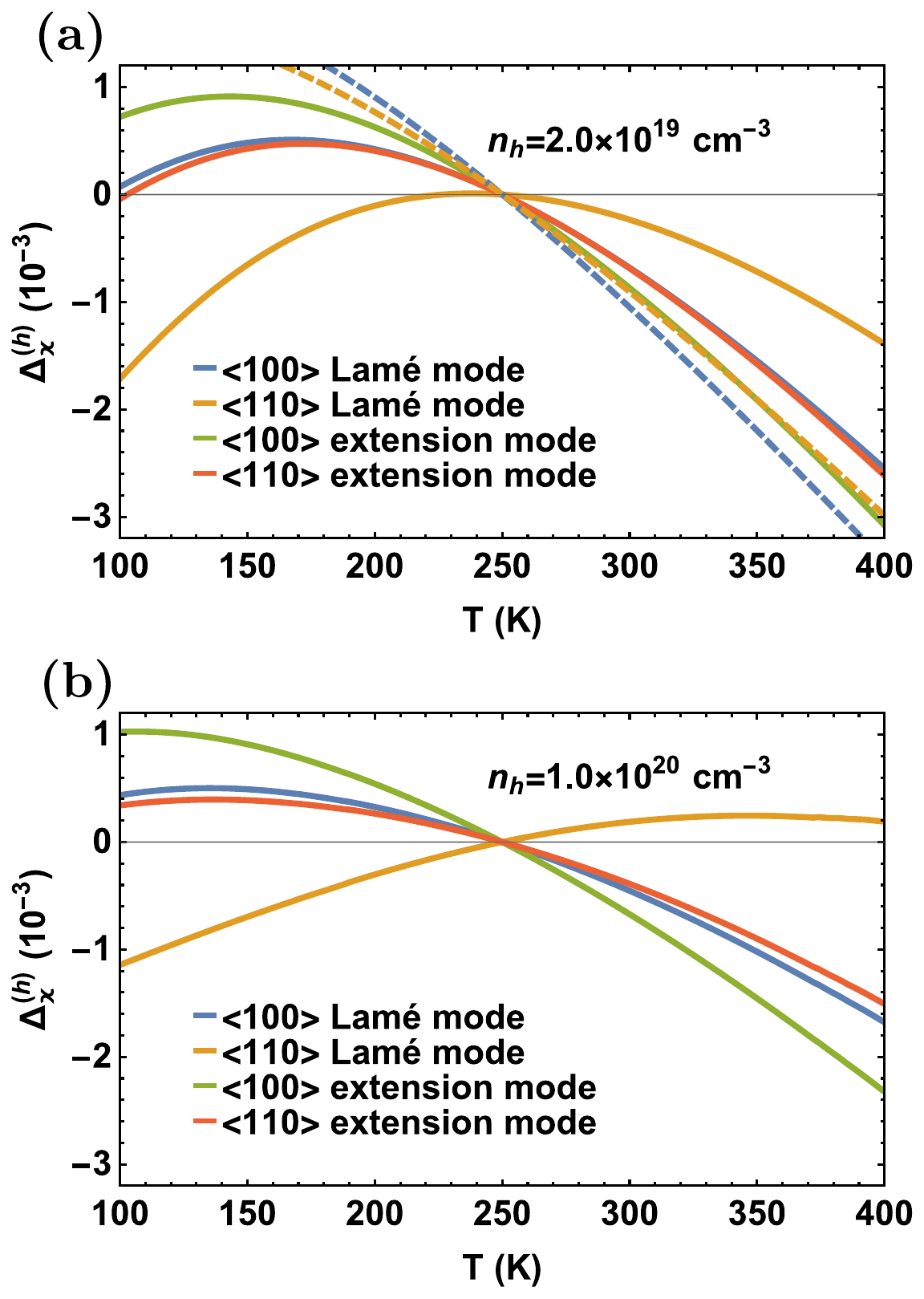}}
	\protect\caption{Temperature dependence of the relative eigenfrequency shifts $\Delta_\ka^{(\mathrm{h})}=[\omega_{\ka}(T)-\omega_{\ka}(T_{0})]/\omega_{\ka}(T_{0})$ at $T_{0}=250$ K for several eigenmodes in Si-based resonators. The values of $\Delta_\ka^\mathrm{(h)}$ are a sum of the contributions from the intrinsic nonlinearity of the elastic constants \cite{varshni1970temperature} and from the coupling to holes. The plots (a) and (b) refer to the hole densities $\nh=2.0\times10^{19}$ cm$^{-3}$ and $\nh=1.0\times10^{20}$ cm$^{-3}$. The solid curves are calculated using the three-band model. The dashed lines in panel (a) show the result of the two-band approximation. }
	\label{fig:Si_three_band_linear}
\end{figure}

Figure \ref{fig:Si_three_band_linear} shows the scaled temperature dependence of the eigenfrequencies for several modes of $p$-doped Si resonators.
We recall that the parameter $\Delta_\ka^{(\mathrm{h})}$, which is defined in Eq.~(\ref{eq:relative_shift}), shows the relative frequency change compared to the frequency value at a certain temperature $T_0$, which we chose to be 250~K. The presented frequency change includes both the frequency shift due to the intrinsic nonlinearity of the crystal (we used the data given in \cite{varshni1970temperature}) and the doping-induced corrections to the second-order elastic constants that we found. The incorporation of the effect of intrinsic nonlinearity is another distinction from the analysis in \cite{Khan1985}. This inclusion is necessary for a comparison of the theory and experiment.

A remarkable feature of the plots in Fig.~\ref{fig:Si_three_band_linear} is that the temperature dependence of the frequencies of all modes we studied is nonmonotonic. This is a result of the competition between the intrinsic and doping-induced frequency shifts. In an undoped crystal, the elasticity parameters, and thus $\Delta_\ka^\mathrm{(h)}$, decrease with temperature in the shown temperature range. In contrast, the coupling to holes leads to an increase of $\Delta_\ka^\mathrm{(h)}$ due to the way this parameter is constructed: the doping-induced correction to the elasticity parameters is negative, but it decreases in the absolute value with the increasing temperature. The possibility of the decrease of the magnitude of the doping-induced correction with rising temperature is clear already from the formal estimate of the coupling effect as $\lVert\widehat\Lambda^{(2)}\rVert\propto \mathcal{D}^2/E_\mathrm{kin}$, since $E_\mathrm{kin} \sim\mathrm{max}(\mu_0,k_BT)$. For higher temperatures, where the effect of the coupling is weak, $\Delta_\ka^\mathrm{(h)}$ decreases with the increasing $T$, as it does in an undoped crystal.

For the density $\nh = 2.0\times 10^{19}\,\mathrm{cm}^{-3}$ in Fig.~\ref{fig:Si_three_band_linear}~(a) the ratio $\mu_0/k_BT$ decreases with the increasing temperature from $3.1$ at $T=100$~K to $0.33$ for $T=250$~K to $-0.67$ for $T=400$~K. For the same temperature values, at the density $\nh = 1.0\times 10^{20}\,\mathrm{cm}^{-3}$ in Fig.~\ref{fig:Si_three_band_linear}~(b) the ratio $\mu_0/k_BT$ decreases from $8.0$ to $2.9$ to $1.4$. It is this difference between the values of $\mu_0/k_BT$ that lies at the root of the difference in the temperature dependence of the eigenfrequencies in the panels (a) and (b) in Fig.~\ref{fig:Si_three_band_linear}. Overall, the coupling-induced frequency change is a nontrivial function of the hole density. More details on the origin of this effect are provided in Section~\ref{subsec:nonlinearity_3band} and in Appendix~\ref{sec:c44_T_depen}.

It is instructive to compare the three-bands results with the ones obtained in the two-band approximation. The two-band results for the $\braket{100}$ and $\braket{110}$ Lam\'e modes are shown with dashed blue and yellow lines in Fig. \ref{fig:Si_three_band_linear}~(a). The significant difference between the two- and three-band results indicates the important role of the split-off band. 


\subsection{Comparison with the experiment}
\label{subsec:experiment}

\begin{figure}[htp!]
	\centerline{\includegraphics[width=0.9\columnwidth]{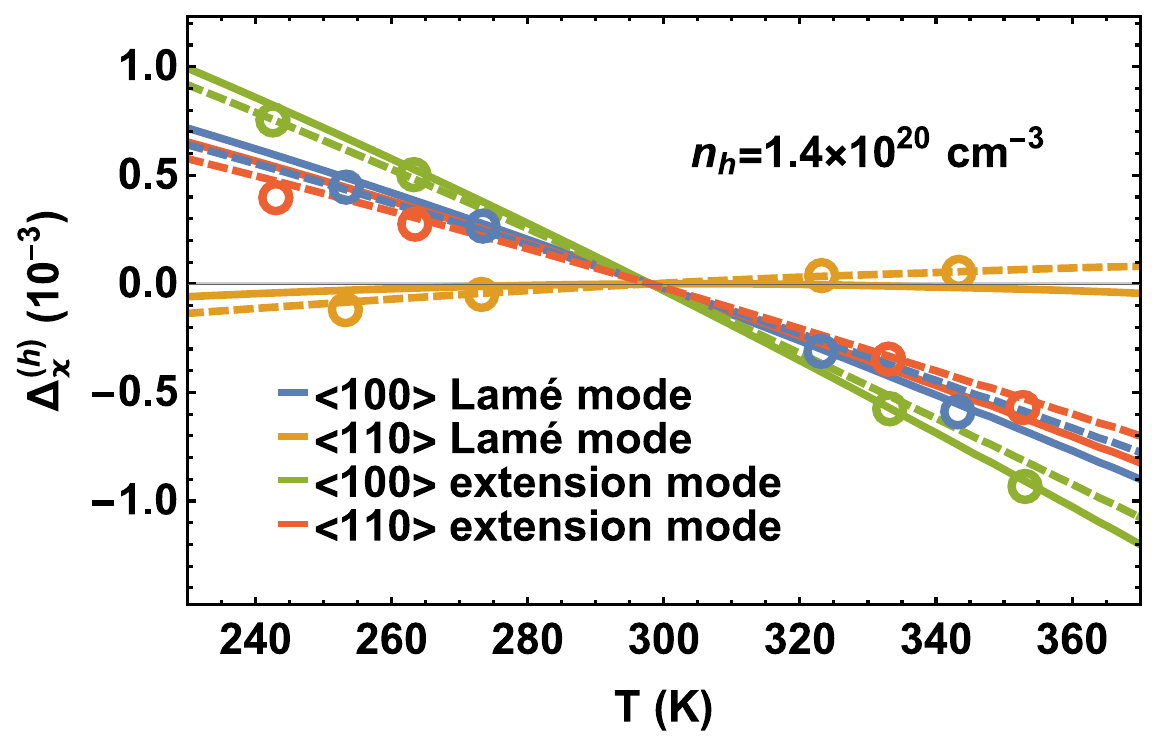}}
	\protect\caption{Temperature dependence of the relative linear frequency shifts $\Delta_\ka^{(\mathrm{h})}=[\omega_{\ka}(T)-\omega_{\ka}(T_{0})]/\omega_{\ka}(T_{0})$ for different modes in Si MEMS at $\nh=1.4\times10^{20}$ cm$^{-3}$ and $T_{0}=298$~K. The solid curves show the results that account only for the $T$ dependence of the elasticity constants, as in Fig.~\ref{fig:Si_three_band_linear}. The dashed curves show the results that additionally take into account thermal expansion using the interpolating expression in Ref.~\cite{Bourgeois1997}. The empty circles are the experimental data \cite{Ng2015} for the corresponding modes.}
	\label{fig:Si_exp_compared}
\end{figure}

In Fig. \ref{fig:Si_exp_compared}, we compare the theory with the experimental results \cite{Ng2015} on the mode eigenfrequencies in Si MEMS for $\nh = 1.4\times 10^{20}~\mathrm{cm}^{-3}$. The theoretical calculations are done for the three-band model. The hole-hole interaction is disregarded, which is justified for such high hole density. Overall, the theory and the measured temperature dependence of the mode eigenfrequencies are in good agreement.

A plausible cause of the small discrepancy between the theory and the experiment is the dependence of thermal expansion on doping. To illustrate the effect of thermal expansion, in Fig.~\ref{fig:Si_exp_compared} we present two theoretical results for each mode. The first refers to the approximation in which the temperature dependence of the eigenfrequencies is due to the temperature dependence of the elastic constants only. The second takes into account thermal expansion of the resonator. To describe the latter effect we employed the results of Refs.~\cite{Lyon1977,Bourgeois1997}, following Ref.~\cite{Ng2015}. These results refer to high-purity Silicon. However, $p$-doping should affect thermal expansion, as seen already from the linear in $\hat\ep$ term in the free energy, Eq.~(\ref{eq:first_order}).

An additional argument in support of the effect of $p$-doping on thermal expansion is as follows. Our calculations show that increasing the hole density in Si beyond $1.0\times 10^{20}\,\mathrm{cm}^{-3}$ results in a slight suppression of the temperature dependence of the doping-induced corrections to the parameters of linear elasticity. This is despite the magnitude of these corrections increasing with the increasing $\nh$; for example, $\lVert\widehat\Lambda^{(2)}\rVert\sim \nh \bigl(\mathcal{D}^{2}/\mu_0\bigr)\propto \nh^{1/3}$, for a strongly degenerate hole gas. However, the temperature dependences observed in Ref.~\cite{Ng2015} for $\nh = 1.4\times 10^{20}\,\mathrm{cm}^{-3}$ and $\nh = 1.7\times 10^{20}\,\mathrm{cm}^{-3}$ are very similar. We emphasize that, as seen from Fig.~\ref{fig:Si_exp_compared}, the overall effect of thermal expansion is small.


\subsection{Mode nonlinearity due to the strain-hole coupling}
\label{subsec:nonlinearity_3band}

\begin{figure}[htp!]
	\centerline{\includegraphics[width=1.0\columnwidth]{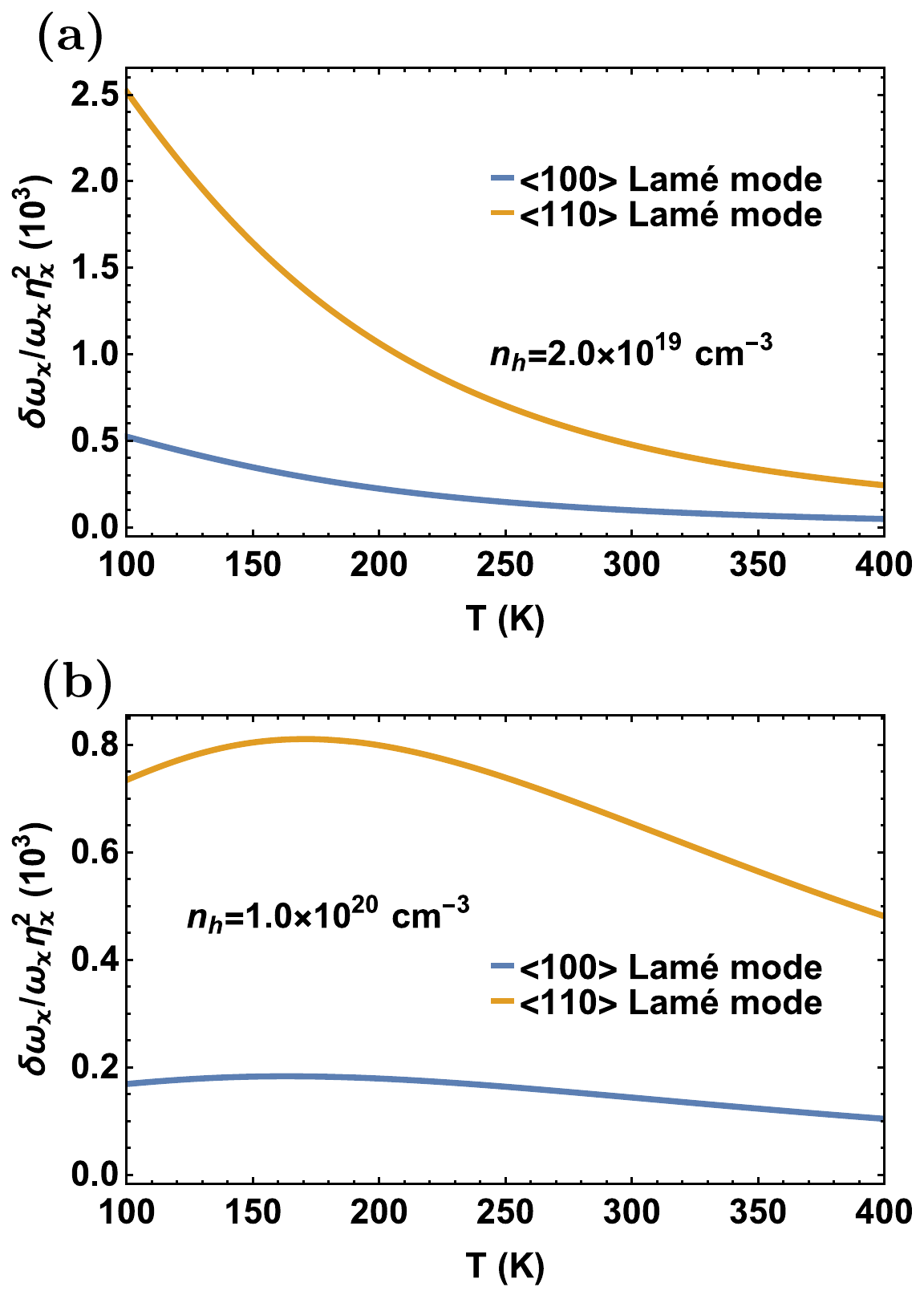}}
	\protect\caption{Temperature dependence of the scaled relative nonlinear frequency shifts $\delta \omega_\ka/\omega_\ka \eta_{\ka}^{2}$ for $\braket{100}$ and $\braket{110}$ Lam\'e modes in single-crystal Si at (a) $\nh=2.0\times10^{19}$ cm$^{-3}$ and (b) $\nh=1.0\times10^{20}$ cm$^{-3}$. Note that the results presented here are calculated through the three-band model as well.}
	\label{fig:Si_three_band_nonlinear_T_depen}
\end{figure}

Figure~\ref{fig:Si_three_band_nonlinear_T_depen} shows the calculated temperature dependence of the scaled nonlinear frequency shift $\aleph_\ka = \delta\omega_\ka/\omega_\ka\eta_\ka^2 $, Eq.~(\ref{eq:scaled_nonlinear_param}), for two Lam\'e modes in $p$-doped Silicon.
The calculations are done using the three-band model and refer to the hole densities $2.0\times10^{19}$~cm$^{-3}$ and $1.0\times10^{20}$~cm$^{-3}$. We note first that the values of $\aleph_\ka$ in Fig.~\ref{fig:Si_three_band_nonlinear_T_depen}~(a) are larger by a factor $>10$ than those for GaAs. This is a consequence of the small band splitting in Silicon and the associated absence of partial compensation of the contributions from light and heavy holes away from strong degeneracy. The calculation in the two-band approximation with the Si parameters gives a much smaller $\aleph_\ka$ than in the three-band model. 

The parameter $\aleph_\ka$ shows a very different behavior as a function of temperature depending on the hole density. For $\nh=2.0\times10^{19}$~cm$^{-3}$, Fig.~\ref{fig:Si_three_band_nonlinear_T_depen}~(a), $\aleph_\ka$ monotonically decreases with the increasing temperature, whereas for $\nh=1.0\times10^{20}$~cm$^{-3}$, Fig.~\ref{fig:Si_three_band_nonlinear_T_depen}~(b), it displays a maximum as a function of $T$. This is different from the temperature dependence of the eigenfrequencies in Fig.~\ref{fig:Si_three_band_linear}. We relate this difference to the following: $\aleph_\ka$ has contributions from the terms in the free energy that contain integrals over $\kb$ of higher-order derivatives of $f_\nu$ over $\mu_0$. For $\nh=2.0\times10^{19}$~cm$^{-3}$, where the hole gas is not strongly degenerate and essentially nondegenerate in the split-off band, these higher-order derivatives make a comparatively large contribution to $\aleph_\ka$. This contribution falls off with the increasing temperature, as explained at the end of Section~\ref{sec:nonlinearity}, and so does $\aleph_\ka$, too. In contrast, for $\nh=1.0\times10^{20}$~cm$^{-3}$, the hole gas is strongly degenerate for $T=100$~K, and then the terms with the higher-order derivatives of $f_\nu$, which have very sharp peaks/dips with zero area, begin contributing as these peaks/dips get broadened with the increasing $T$, leading to an initial increase of $\aleph_\ka$, followed by the decrease for still larger $T$. 

\begin{figure}[htp!]
	\centerline{\includegraphics[width=1.0\columnwidth]{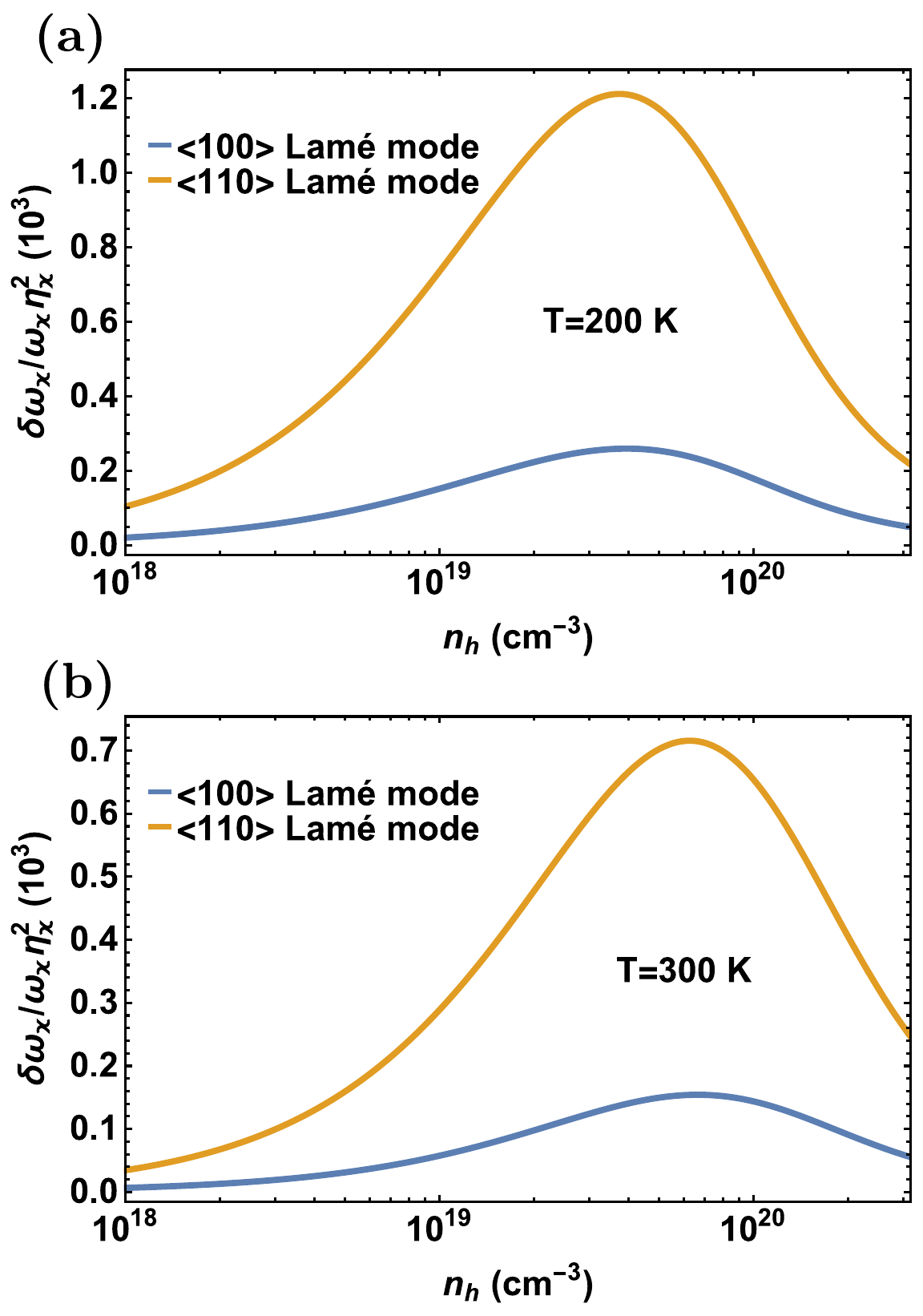}}
	\protect\caption{Hole density dependence of the scaled relative nonlinear frequency shifts $\delta \omega_\ka/\omega_\ka \eta_{\ka}^{2}$ for $\braket{100}$ and $\braket{110}$ Lam\'e modes in single-crystal Si at (a) $T=200$ K and (b) $T=300$ K. The results here are obtained using the three-band calculations.}
	\label{fig:Si_three_band_nonlinear_n_depen}
\end{figure}

The above arguments are corroborated by 
Fig. \ref{fig:Si_three_band_nonlinear_n_depen}, which shows the dependence of $\aleph_\ka$ on the hole density for $\braket{100}$ and $\braket{110}$ Lam\'e modes at $T=200$~K and $T=300$~K.
The increase of $\aleph_\ka$ with $\nh$ for low densities is a consequence of the increase of the effect of the coupling for a nondegenerate hole gas. On the other hand, for $\mu_0\gg k_BT$, where the gas is strongly degenerate, $\aleph_\ka$ is decreasing with the increasing $\nh$. This is because the major contribution to $\aleph_\ka$ in this density range comes from the term in the free energy given by the integral over $\kb$ of the sum $\sum_\nu f_\nu E_\nu\4(\kb)$, cf. Eq.~(\ref{eq:F_fourth}). Since $\sum_\nu E_\nu\4(\kb) =0$, the contribution to this sum comes from the range of $\kb$ where some bands $\nu$ are partially occupied. Furthermore, since $ E_\nu\4(\kb)\propto (\mathcal{D}\lVert\hat\ep\rVert/\mu_0)^4$ for the corresponding energies and $\mu_0\propto \nh^{2/3}$, $\aleph_\ka$ falls off with the increasing $\nh$. The crossover from the increase to decrease with the increasing $\nh$, which is due to the onset of strong degeneracy of the hole gas, shifts to higher $\nh$ with the increasing temperature.


\section{Effect of hole-strain coupling on mode damping}
\label{sec:Akhiezer}

In this section we provide a brief discussion of another effect of the coupling to holes, the effect of this coupling on the decay rates of the low-frequency eigenmodes. Here the major mechanisms include direct decay due to hole scattering off the eigenmodes and changes in decay resulting from their coupling to higher-frequency phonons. These changes consist of the modifications to the very coupling parameters caused by doping, as well as changes in the decay rates of thermal phonons arising from their interactions with holes and scattering off dopants. It is important to note that the very presence of dopants does not directly lead to the decay of the low-frequency eigenmodes, although it changes their eigenfrequencies.

To the leading order in the hole-strain coupling, the rate of the direct decay due to scattering of holes off the low-frequency eigenmode is determined by the probabilities of transitions where a hole goes from its initial to a final state $\ket{\psi_\mathrm{i}}\to \ket{\psi_\mathrm{f}}$ and a vibrational quantum of the mode is absorbed. The probabilities of such transitions are proportional to the difference in the populations of the states $\ket{\psi_\mathrm{i}}$ and $\ket{\psi_\mathrm{f}}$, i.e., it is $\propto \hbar\omega_\ka f'_\nu \propto (\hbar\omega_\ka/E_\mathrm{kin})$, where $E_\mathrm{kin} = \max(\mu_0,k_BT)$ is the typical hole energy, with $E_\mathrm{kin}\gg \hbar\omega_\ka$. This strong inequality reduces the scattering rate.

Other factors that come into play are related to the dynamics of holes. The hole mean free path in a highly doped Si micromechanical resonator is in the range of $10^{-2}\,\mu\mathrm{m}$. It is much smaller than the length scale on which the strain created by low-frequency eigenmodes varies. Note that this length scale is given by the resonator size $L\sim 10^2~\mu\mathrm{m}$. Another important factor is that the hole relaxation time, $\tau_\mathrm{h}$, is much smaller than the reciprocal frequencies of the modes under consideration, i.e., $\omega_\ka\tau_\mathrm{h}\ll 1$. For highly doped Si, $\tau_\mathrm{h}$ is on the order of $10^{-13}$--$10^{-14}$~s at temperatures near room temperature. Therefore, when estimating the mode decay rate due to transition $\ket{\psi_\mathrm{i}}\to \ket{\psi_\mathrm{f}}$, where holes scatter off the mode, the standard $\delta$-function enforcing energy conservation, $E_\mathrm{f}-E_\mathrm{i}=\hbar\omega_\ka$, can be replaced by $\tau_\mathrm{h}/\hbar$ in the expression for the scattering rate.

The coupling Hamiltonian can be estimated as $\mathcal{H}_i\sim \mathcal{D} (\hbar/2M\omega_\ka)^{1/2}\lVert \ep_\ka(\rb)\rVert$, cf. Section~\ref{subsec:estimate}. With this estimate, and using the appropriately modified Fermi golden rule, we obtain the following estimate for the mode decay rate $\Gamma_\ka^{(\mathrm{h})}$ due to direct hole scattering:
\begin{align}
\label{eq:decay_hole_direct}
\Gamma_\ka^{(\mathrm{h})}\sim \frac{1}{\rho} \frac{\mathcal{D}^2}{E_\mathrm{kin}}L^{-2}\tau_\mathrm{h} \nh,
\end{align}
where we used the relation $M=\rho V$. For room temperature and the parameters from previous estimates (cf. Section~\ref{subsec:estimate}), this yields $ \Gamma_\ka^{(\mathrm{h})}\sim 1\,\mathrm{s}^{-1}$ for $L\sim 100\,\mu\mathrm{m}$. For a nondegenerate hole gas, $\Gamma_\ka^{(\mathrm{h})}$ falls off with the increasing temperature because $E_\mathrm{kin}=k_BT$, and at the same time, the hole relaxation time $\tau_\mathrm{h}$ decreases.

We now discuss the decay mechanism associated with the scattering of high-frequency (thermal) phonons off the low-frequency eigenmode of the resonator. In MEMS, this type of scattering primarily arises from the cubic anharmonicity, i.e., the terms $\propto \hat\ep^3$ in the free energy. In pure and $n$-doped Si, the coefficients of different components of the rank-6 tensor $\hat\ep^3$ (i.e., the third-order elastic constants $c_{ijk}$) have different signs \cite{Hall1967,Philip1983}. As mentioned earlier, doping with acceptors introduces significant contributions $c_{ijk}^{(\mathrm{h})}$ to these coefficients, which can be comparable to the intrinsic values and may either increase or decrease them (these corrections also have different signs in $p$-doped Si; in particular, within the two-band approximation we found that $c_{111}^{(\mathrm{h})}=c_{123}^{(\mathrm{h})}=-2c_{112}^{(\mathrm{h})}$ and $c_{144}^{(\mathrm{h})}=-2c_{155}^{(\mathrm{h})}$). The modification of the third-order elastic constants due to the $p$-type doping can either increase or decrease the decay rate resulting from phonon scattering, depending on which phonon branches are  more strongly coupled to the low-frequency eigenmode.

Three mechanisms of phonon-induced decay of the low-frequency MEMS modes are typically considered: the Landau-Rumer mechanism, thermoelastic relaxation, and Akhiezer damping, cf. Ref.~\cite{Bachtold2022a}. The Landau-Rumer mechanism applies when the relaxation time of the high-frequency phonons, $\tau_\mathrm{ph}$, exceeds $\omega_\ka^{-1}$. In pure Si crystals at room temperature, $\tau_\mathrm{ph} \sim 10^{-10}$~s \cite{Regner2013}. This is consistent with the estimate \cite{lifshitz1981physical,Gurevich1988} 
\[\tau_{\text{ph}}^{-1}\sim k_{B}T/m_0v_{s}a_{0},\]
where $m_0$ and $a_0$ are the atomic mass and the lattice constant of Si, respectively. The value of $\tau_\mathrm{ph}$ somewhat varies between different phonon branches, cf. Ref.~\cite{Regner2013}; however, since $\tau_\mathrm{ph}^{-1} \gg \omega_\ka$, the Landau-Rumer mechanism is not relevant.

The theory of thermoelastic relaxation and Akhiezer damping takes into account the fast relaxation of high-frequency phonons. Thermoelastic relaxation is important when $\omega_\ka^{-1}$ is comparable to the time that required for heat to propagate across the region where strain is nonuniform, assuming that the phonon thermalization time is much smaller than $\omega_\ka^{-1}$. For the strain from the low-frequency eigenmodes considered here, the heat diffusion time is $\sim L^2/v_s^2\tau_\mathrm{ph}\gg \omega_\ka^{-1}$. Therefore, the major mode decay mechanism is the Akhiezer damping. Note that we have not discussed the damping of the low-frequency MEMS mode due to its coupling to two-level systems, as this mechanism likely plays a minor role at room temperature.

The Akhiezer damping of low-frequency MEMS eigenmodes has been considered in a number of theoretical papers, cf. \cite{Kiselev2008,Kunal2014,Atalaya2016,Iyer2016,Hamoumi2018,Bachtold2022a}. The temperature dependence of this damping in Si MEMS was studied experimentally in \cite{Rodriguez2019}. In the regime where $\omega_\ka\tau_\mathrm{ph} \ll 1$, the decay rate can be written as
\begin{align}
\label{eq:Akhiezer}
\Gamma_\ka^\mathrm{(Akh)} = a^\mathrm{(Akh)} C \gamma_\mathrm{cpl}^2 \omega_\ka\tau_\mathrm{ph}.
\end{align}
This expression disregards the difference between the relaxation rates of different high-frequency phonons; $C$ is the specific heat per unit mass, and $\gamma_\mathrm{cpl}^2$ is the characteristic variance of the scaled coupling parameters between the considered low-frequency mode and different high-frequency phonons. It can be thought of as the variance of the generalized Gr\"uneisen parameters \cite{Iyer2016,Atalaya2016,Bachtold2022a} and is affected by doping, as discussed earlier. The parameter $a^\mathrm{(Akh)}$ is model-dependent; for a simple model of coupling to acoustic phonons, it is given by $a^\mathrm{(Akh)} = \omega_\ka T/3v_s^2$ \cite{Woodruff1961}. As an estimate, for $\omega_\ka\sim 10^{8}~\mathrm{s}^{-1}$, $\tau_\mathrm{ph}\sim 10^{-11}~\mathrm{s}$, and typical parameter values for Si at room temperature \cite{madelung2004semiconductor}, with $\gamma_\mathrm{cpl}\sim 0.4$ \cite{gauster1971lowtemp}, we obtain $\Gamma_\ka^{(\mathrm{Akh})} \sim 50~\mathrm{s}^{-1}$, which is indeed much larger than $\Gamma_\ka^{(\mathrm{h})}$.

The coupling of high-frequency phonons to holes and acceptors opens additional scattering channels. As a result, increasing the doping level leads to a decrease in the phonon relaxation time $\tau_\mathrm{ph}$, see Refs. \cite{Asheghi2002} and \cite{dongre2020combined} and the papers cited therein. In turn, as seen from Eq.~(\ref{eq:Akhiezer}), this generally results in a decrease in the decay rate $\Gamma_\ka^\mathrm{(Akh)}$ with the increasing doping. However, as mentioned earlier, the doping-induced corrections to the third-order elastic constants may enhance or reduce the coupling between the considered MEMS modes and the relevant high-frequency phonons, thus affecting $\gamma_\mathrm{cpl}^2$. 

Taken together, these arguments indicate that, quite unexpectedly, the decay rate may either increase or decrease with the increasing level of doping. A detailed analysis of the effect is beyond the scope of this paper and will be done separately.


\section{Conclusions}
\label{sec:conclusions}

The results of this paper show a strong effect of the hole-strain coupling on the eigenmodes of semiconductor-based nano- and micromechanical resonators. The effect is rooted in the large values of the deformation potential parameters in typical semiconductors. We consider the range of densities and temperatures where it is sufficient to take into account only two or three hole energy bands. The two-band analysis applies to semiconductors with a comparatively large spin-orbit splitting in the valence band, as in Ge or GaAs, for example. If the spin-orbit splitting is small, as in Si, it is necessary to take the split-off band into account for the hole densities and temperatures of interest.

Where the two-band approximation applies, we provide explicit expressions for the doping-induced corrections to the linear elasticity that account for band warping, and we analyze the corrections to the quartic nonlinearity of the eigenmodes. To describe the effect of the coupling to holes where the split-off band needs to be taken into account, we develop an efficient numerical algorithm. We also briefly outline two effects of the coupling on the decay of low-frequency eigenmodes. One is the direct hole scattering off the modes, which we show to be a weak effect. The other is the change of the decay rates of thermal phonons and their coupling to low-frequency eigenmodes. Our estimate shows that, depending on the intrinsic anharmonicity parameters of the crystal, this change can lead to an increase or, unexpectedly, a decrease of the decay rates of the low-frequency eigenmodes. However, the full quantitative analysis of the decay rate is beyond the scope of this paper.

Our results reveal a nontrivial interplay between the band structure and the degeneracy of the hole gas. In the two-band model, we illustrate this interplay using GaAs resonators, which are an important class of relevant systems. Here, in a broad range where the ratio of the chemical potential to the temperature varies from 4.4 to 0.27, the coupling-induced change of the eigenmode frequencies remains small. This is primarily due to the strain-induced corrections to the light- and heavy-hole band energies having opposite signs and thus effectively compensating each other when integrated over the bands 

The compensation is much less efficient for the quartic (Duffing) nonlinearity parameter of the modes, leading to a nonmonotonic temperature dependence of this parameter. We present this dependence for the Lam\'e modes, which are often studied in experiments. The difference between the linear and nonlinear effects arises, because the expression for the nonlinearity parameter includes higher-order derivatives of the densities of states weighted with the coupling which, for different bands, vary with energy more strongly than the weighted densities of states themselves.

The changes of the mode parameters are significantly more pronounced for the hole densities and temperatures where the split-off band is partially occupied. The mode eigenfrequencies are changed much stronger than in the two-band approximation, where this band is disregarded. They display a distinctly nonmonotonic temperature dependence. The Duffing nonlinearity parameter is significantly larger. 


Interestingly, the effect of doping on the nonlinearity depends nonmonotonically on the hole density $\nh$. For low densities, the effect increases with the increasing $\nh$ as there are more holes available to couple to. However, for higher densities, as the hole gas becomes strongly degenerate, the energies of the holes that can change their states due to strain become larger, since these energies are determined by the chemical potential. Respectively, the effect of the perturbation from the strain becomes weaker. 

In terms of applications, particularly those that require frequency stability, it is important that doping modifies the temperature dependence of the frequencies of different modes in different ways. Therefore one can detect a temperature change by monitoring the frequency of one mode, and then compensate it so as to adjust the frequency of the other mode to a desired value. In terms of suppressing temperature-related frequency fluctuations, it is also advantageous that there emerge plateaus in the temperature dependence of the eigenfrequencies of some modes. The sensitivity and nontrivial dependence of the eigenfrequency change and the mode nonlinearity on the hole density allow one to find an optimal doping level in the desired temperature range.

The results of the paper are in good agreement with the experimental data on the temperature dependence of the eigenmode frequencies in $p$-doped Silicon microresonators. They explain why this dependence is different for different modes and provide means for predicting this dependence and the dependence of the mode nonlinearity on the hole density and temperature. More broadly, the results show that nano- and micromechanics reveal new aspects of the electron-phonon coupling, which are of interest both for fundamental and applied studies.

\acknowledgments

This project was sponsored by the Defense Advanced Research Projects Agency (DARPA) under cooperative agreement HR0011-23-2-0004. The content of this paper does not reflect the position or the policy of the US Government, and no official endorsement should be inferred.


\appendix

\section{The free energy expansion}
\label{sec:free_energy}

We calculate the expansion terms of the free energy $\Delta{\cal F}\un$ in two steps: first we find the the strain-induced corrections to the hole energy $E_\nu\un(\kb,\hat\ep)$ using the full Hamiltonian $\hat H(\kb,\hat\ep)$, and then we sum these corrections over the bands $\nu$ and integrate over $\kb$, with the proper weight. In this section we give the expressions for the weighting factors. They are obtained by expanding the grand potential $\Omega_\nu(\rb)$ in $\mu\un(\hat\ep) - E_\nu\un(\kb,\hat\ep)$ and taking into account that the hole density is not perturbed by the strain. To shorten the notation, we introduce an operator
\begin{align}
\label{eq:sum_operator}
\hat \SS = \sum_\nu \int d^3\kb/(2\pi)^3
\end{align} 
that describes summation over the bands and integration over $\kb$.

We have provided the expressions for $\mu\1$ and $\Delta\mathcal{F}\1$ in Eq.~(\ref{eq:first_order}). To the second order, we have the correction to the chemical potential
\begin{align}
\label{eq:mu_second}
	\mu\2 
	=&(\partial\nh/\partial\mu\0)^{-1}\frac{\partial}{\partial \mu\0} \hat \SS \left [2f_\nu E_\nu\2 \right. \nonumber
	\\
	&\left. - f'_\nu(\mu\1 - E_\nu\1)^2\right],
\end{align}
where, as in the main text, we adopt the notation $f_\nu\equiv f_\nu(E_\nu\0, \mu\0)$ and $E_\nu\un\equiv E_\nu\un(\kb,\hat\ep)$.

The second-order term in the free energy is
\begin{align}
\label{eq:F_second}
\Delta\mathcal{F}\2 = \hat\SS\left[2f_\nu E_\nu\2 - f'_\nu (\mu\1 - E_\nu\1)^2\right],
\end{align}
where $f'_\nu\equiv \partial f_\nu(E_\nu\0,\mu\0)/\partial\mu\0$; in the main text and in what follows, the number of primes after $f_\nu$ denotes the order of the derivative of $f_\nu(E_\nu\0,\mu\0)$ over $\mu\0$.

Similarly we find 
\begin{align}
\label{eq:mu_third}
	&\mu\3 
	=(\partial\nh/\partial\mu\0)^{-1}\frac{\partial}{\partial \mu\0} \hat \SS \left [2f_\nu E_\nu\3 \right. \nonumber\\
	&\left. - 2f'_\nu(\mu\1 - E_\nu\1)(\mu\2 - E_\nu\2)
	 -\frac{1}{3}f''_\nu (\mu\1 - E_\nu\1)^3 \right]
\end{align}
and
\begin{align}
\label{eq:F_third}
&\Delta\mathcal{F}\3 = \hat\SS\left[2f_\nu E_\nu\3 - 2 f'_\nu (\mu\1 - E_\nu\1)(\mu\2 - E_\nu\2)\right. \nonumber\\
&\left. - \frac{1}{3} f''_\nu(\mu\1 - E_\nu\1)^3
\right ].
\end{align}

To find the amplitude dependence of the eigenmode frequency, we also need the fourth-order term in the free energy:
\begin{align}
\label{eq:F_fourth}
 &\Delta\mathcal{F}\4 = \hat\SS\left\{2f_\nu E_\nu\4
 - f'_\nu \left[(\mu\2 - E_\nu\2)^2 \right.\right.\nonumber\\ 
&\left.\left. +2(\mu\1 - E_\nu\1)(\mu\3 - E_\nu\3)\right]
 \right. \nonumber\\
&\left. - f''_\nu(\mu\1 - E_\nu\1)^2 (\mu\2 - E_\nu\2)
-\frac{1}{12}f'''_\nu(\mu\1 - E_\nu\1)^4
\right\}.
\end{align}

The calculation in the two-band approximation is simplified by the fact that $\mu\1 = a\tr \hat \ep$ is equal to the isotropic part of $E_\nu\1$. Therefore, $\mu\3$ drops out from $\Delta\mathcal{F}\4$, and it is sufficient to find only $\mu\2$.


\section{Cancellation of the long-wavelength divergence in the free energy expansion}
\label{sec:divergence}

The individual terms in the expression for the correction to the free energy of the fourth-order in strain $\Delta\mathcal{F}\4$, Eq.~(\ref{eq:F_fourth}), diverge for $\kb \to {\bf 0}$. However, we show in this section that the overall expression remains finite in the two-band approximation. In this approximation we have from Eq.~(\ref{eq:strained_dispersion}) $E_{\nu}^{(n)}(\kb,\hat{\ep})\propto 1/k^{2(n-1)}$ for $\kb\rightarrow{\bf 0}$. Therefore, the correction to the chemical potential $\mu\2$, which is given by Eq.~(\ref{eq:mu_second}), does not diverge. The integrals over $\kb$ of the terms that contain $\sum_\nu f_{\nu}^{\prime\prime}\tilde{E}_{\nu}\1{}^2\mu\2$, $\sum_\nu f_{\nu}^{\prime}\tilde{E}_{\nu}\2\mu\2$, $\sum_\nu f_{\nu}^{\prime\prime}\tilde{E}_{\nu}\1{}^2\tilde{E}_{\nu}\2$, and $\sum_\nu f_{\nu}^{\prime\prime\prime}\tilde{E}_{\nu}\1{}^4$ do not diverge either. Here $\tilde{E}_\nu^{(n)} \equiv E_\nu^{(n)}$ for $n\geq2$ while $\tilde{E}_\nu\1 \equiv E_\nu\1-a\tr\hat{\ep}$; we also use the fact that $\tilde{E}_0^{(n)} = -\tilde{E}_1^{(n)}$. 
As a result, it is sufficient to check the convergence of the contribution to $\Delta\mathcal{F}\4$ of the remaining sum of the singular terms,
\begin{align}
\label{eq:sum_diverging}
S_\mathrm{sing}=\sum_\nu \left[f_{\nu}\tilde{E}_{\nu}\4 -f_{\nu}^{\prime}\left(\tilde{E}_{\nu}\1 \tilde{E}_{\nu}\3 +\frac{1}{2}\tilde{E}_\nu\2{}^2\right)\right].
\end{align}

The term $\tilde{E}_\nu\4$ in $S_\mathrm{sing}$ is $\propto k^{-6}$ for $k\to 0$; this is the strongest divergence in the expression. However, one has to take into account that the contributions $\tilde{E}_\nu\un$ from the light and heavy holes to the sums over $\nu=0,1$ have opposite signs, cf. Eq.~(\ref{eq:strained_dispersion}). As a result, if we replace $f_\nu$ with its $\nu$-independent value at zero energy, $f(0,\mu\0) = [1+\exp(-\mu\0/k_BT)]^{-1}$, the terms $\propto k^{-6}$ cancel each other. If one further uses the expansion $f_\nu\equiv f_\nu(E_\nu\0,\mu\0) \approx f(0,\mu\0) - f'(0,\mu\0)E_\nu\0(\kb)$ and, again taking into account that $\tilde{E}_0^{(n)} = -\tilde{E}_1^{(n)}$, one obtains for $k\to 0$:
\[S_\mathrm{sing} \approx -2f'(0,\mu\0)\left[\mathcal{E}(\kb) \tilde{E}_0\4 + \tilde{E}_0\1\tilde{E}_0\3 +\frac{1}{2}\tilde{E}_0\2{}^2\right].\]
A direct calculation based on Eq.~(\ref{eq:strained_dispersion}) shows that the terms $\propto k^{-4}$ and $k^{-2}$ drop out of $S_\mathrm{sing}$. The remaining term does not give a diverging contribution to $\Delta\mathcal{F}\4$.

These arguments are corroborated by the estimate of the contribution to $\Delta\mathcal{F}$ from the range of small $k<k_\ep$, where the strain-induced term in the hole energy exceeds the ``bare'' energy $E_\nu\0(\kb)$. From Eq.~(\ref{eq:strained_dispersion}), $k_\ep \sim [(b/B)\|\hat\ep\|]^{1/2}$. For such $k$, the hole energy is small compared to the thermal energy, $E_\nu(\kb) \ll k_BT$. Then, for $k<k_\ep$, we can expand the integrand in Eq.~(\ref{eq:Omega_nu}) for $\Omega_\nu(\rb)$ in $\xi_\nu=(-1)^\nu (\mathcal{E}^2 + \Xi\1 + \Xi\2)^{1/2}/k_BT$. The leading-order contribution from the range $k\lesssim k_\ep$ to the free energy is quadratic in $\xi_\nu$ and comes with the weight $k_\ep^3$, which implies that it scales as $\|\hat\ep\|^{7/2}$. This is of the same order of magnitude as the contribution from the range of small $k$ to $\Delta\mathcal{F}$ given by Eqs.~(\ref{eq:Lambda_11_MD}) and (\ref{eq:Lambda_12_MD}). Indeed, taking into account that for small $k$, the difference between $f_0$ and $f_1$ is $\propto k^2$, we see that the integrands in these equations are independent of $k$, so the contribution from the range $k\lesssim k_\ep$ to the harmonic part of $\Delta\mathcal{F}$ scales as $\|\hat\ep\|^{7/2}$ too. It is easy to check that the corresponding contribution to $\Delta\mathcal{F}\4$ scales as $\|\hat\ep\|^{11/2}$, in agreement with $S_\mathrm{sing}=\mathcal{O}(k^0)$ for $k\to 0$.


\section{The Luttinger-Kohn-Bir-Pikus Hamiltonian}
\label{sec:three-band}

Here, for completeness, we provide the explicit expression of the three-band Luttinger-Kohn-Bir-Pikus Hamiltonian and the values of the relevant parameters, along with some properties of the strain-induced corrections to the hole energy used in the main text. We use the notations of Ref.~\cite{winkler2003spin}. The Hamiltonian $\hat H(\kb,\hat\ep)$ for a given wave vector $\kb$ and strain $\hat\ep$ is a $6\times 6$ matrix and has the form $\hat H(\kb,\hat\ep) = \hat H\0(\kb) +\hat{H}_{i}\cdot\hat{\ep}$, where 
$\hat{H}\0(\kb)$ is the free-hole term and $\hat{H}_{i}\cdot\hat{\ep}$ is the strain-induced term. The terms have a similar structure, which is dictated by the symmetry.

The free-hole Hamiltonian $\hat{H}\0(\kb)$ can be written as 
\begin{widetext}
\begin{align*}
	\hat{H}\0(\kb)=\begin{pmatrix}
		\left[\hat{H}\0(\kb)\right]_{4\times4} & \left[\hat{H}\0(\kb)\right]_{4\times2}\\
		\left[\hat{H}\0(\kb)\right]_{2\times4} & \left[\hat{H}\0(\kb)\right]_{2\times2}
	\end{pmatrix},
\end{align*}
where the upper left $4\times4$ block describes the dynamics of the light and heavy holes, the lower right $2\times2$ block is associated with the spin split-off band, and the off-diagonal $4\times2$ and $2\times4$ blocks describe the interband coupling. In the explicit form 
\begin{align}
	\left[\hat{H}\0(\kb)\right]_{4\times4}=\frac{\hbar^2}{2m_{0}}\Bigg[\left(\gamma_1 + \frac{5}{2} \gamma_2\right)\kb^2\mathbb{1}_{4\times4} 
	-2 \gamma_2\sum_{i} k_i^2 \hat{J}_i^2-2\gamma_3 \sum_{i\neq j} k_i k_j \hat{J}_i\hat{J}_j
	\Bigg],
\end{align}
\begin{align*}
	\left[\hat{H}\0(\kb)\right]_{2\times2}=\left(\frac{\hbar^2}{2m_{0}}\gamma_{1}\kb^2+\Delta_{0}\right)\mathbb{1}_{2\times2},
\end{align*}
and
\begin{align*}
	\left[\hat{H}\0(\kb)\right]_{4\times2}=\left[\hat{H}\0(\kb)\right]_{2\times4}^{\dagger}=\frac{\hbar^2}{2m_{0}}\begin{pmatrix}
	\sqrt{6}\gamma_{3}k_{z}(k_{x}-ik_{y}) & \sqrt{6}\left[\gamma_{2}(k_{x}^2-k_{y}^2)-2\gamma_{3}ik_{x}k_{y}\right]\\
	\sqrt{2}\gamma_{2}(k^2-3k_{z}^2) & -3\sqrt{2}\gamma_{3}k_{z}(k_{x}-ik_{y})\\
	-3\sqrt{2}\gamma_{3}k_{z}(k_{x}+ik_{y}) & -\sqrt{2}\gamma_{2}(k^2-3k_{z}^2)\\
	-\sqrt{6}\left[\gamma_{2}(k_{x}^2-k_{y}^2)+2\gamma_{3}ik_{x}k_{y}\right] & \sqrt{6}\gamma_{3}k_{z}(k_{x}+ik_{y})
	\end{pmatrix}.
\end{align*}
Here, $i,j$ enumerate the components $x,y,z$ which are chosen along the $\braket{100}$ axes; $m_{0}$ is the free-electron mass; $\gamma_{1,2,3}$ are the Luttinger parameters; $\Delta_0$ is the band splitting due to spin-orbit coupling; $\mathbb{1}_{4\times4}$ and $\mathbb{1}_{2\times2}$ are the $4\times4$ and $2\times2$ identity matrices, respectively; and $\hat{J}_{x,y,z}$ are the angular-momentum matrices for the angular momentum $J=3/2$. The three Luttinger parameters are $\gamma_{1}\approx4.285$, $\gamma_{2}\approx0.339$, and $\gamma_{3}\approx1.446$ for Si, and $\gamma_{1}\approx6.85$, $\gamma_{2}\approx2.10$, and $\gamma_{3}\approx2.90$ for GaAs.


\begin{figure*}
	\centerline{\includegraphics[width=0.9\textwidth]{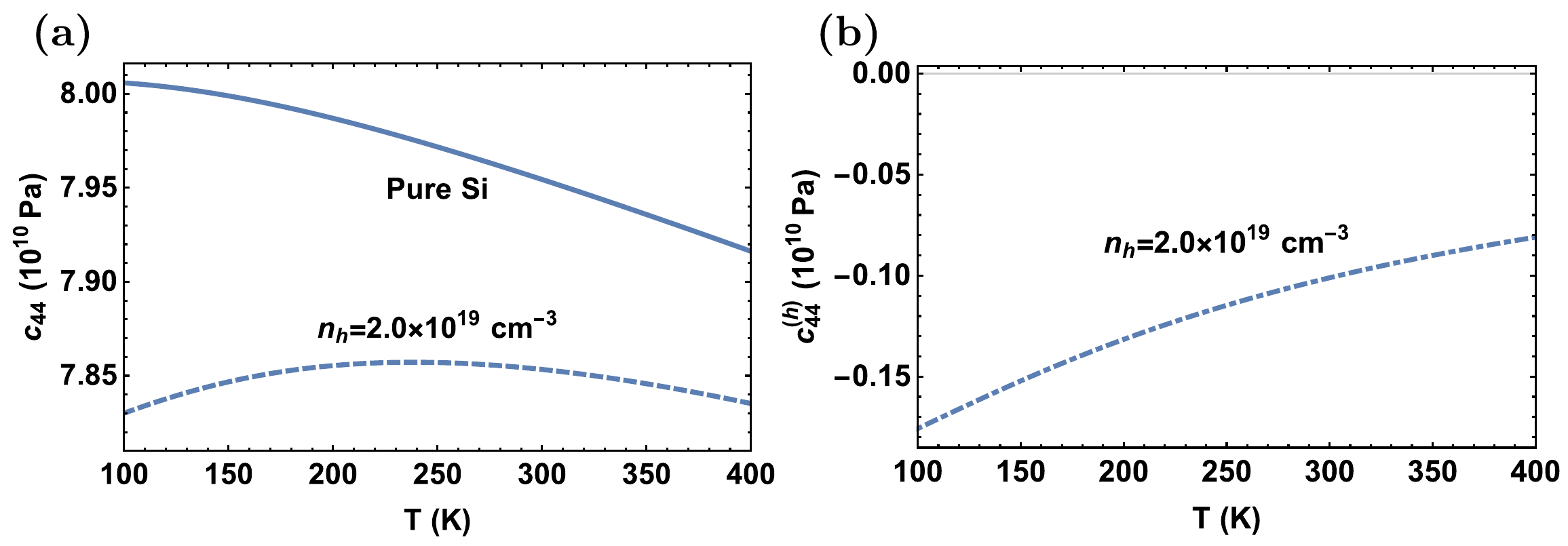}}
	\protect\caption{Temperature dependence of the elastic constant $c_{44}$. Panel (a) shows $c_{44}$ as a function of $T$ for pure (solid curve) and $p$-doped Si with a hole density of $\nh=2.0\times10^{19}$ cm$^{-3}$ (dashed curve). The parameters for pure Si are taken from Ref.~\cite{varshni1970temperature}. Panel (b) shows the temperature dependence of the doping-induced correction $c_{44}^{(\text{h})}$ at $\nh=2.0\times10^{19}$ cm$^{-3}$, which is calculated using the three-band model. The dashed curve in (a) shows the sum of the intrinsic and doping-induced contributions, demonstrating a nonmonotonic temperature dependence of $c_{44}$ in doped Si.}
	\label{fig:Si_c44_T_depen}
\end{figure*}

The coupling Hamiltonian $\hat{H}_{i}\cdot\hat{\ep}$ has the form
\begin{align*}
	\hat{H}_{i}\cdot\hat{\ep}=\begin{pmatrix}
		\left[\hat h_i\right]_{4\times4} & \left[\hat h_i\right]_{4\times2}\\
		\left[\hat h_i\right]_{2\times4} & \left[\hat h_i\right]_{2\times2}
	\end{pmatrix},
\end{align*}
where
\begin{align}
\label{eq:full_perturbation}
	\left[\hat h_i\right]_{4\times4}=-\left(D_{d}-\frac{5}{6}D_u\right)\tr \hat{\ep}\, \mathbb{1}_{4\times4} -
	\frac{2}{3}D_{u}\sum_i \hat{J}_i^2\ep_{ii}-\frac{2}{3}D'_u\sum_{i\neq j}\hat{J}_i \hat{J}_j\ep_{ij},
\quad
	\left[\hat h_i\right]_{2\times2}=-D_{d}\tr\hat{\ep}\,\mathbb{1}_{2\times2},
\end{align}
and
\begin{align}
\label{eq:off_diagonal_H_i}
	\left[\hat h_i\right]_{4\times2}=\left[\hat h_i\right]_{2\times4}^{\dagger}=\begin{pmatrix}
	\sqrt{\frac{2}{3}}D'_{u}(\ep_{zx}-i\ep_{yz}) & \sqrt{\frac{2}{3}}\left[D_{u}(\ep_{xx}-\ep_{yy})-2D'_{u}i\ep_{xy}\right]\\
	\frac{\sqrt{2}}{3}D_{u}\left[\tr\hat{\ep}\,-3\ep_{zz}\right] & -\sqrt{2}D'_{u}(\ep_{zx}-i\ep_{yz})\\
	-\sqrt{2}D'_{u}(\ep_{zx}+i\ep_{yz}) & -\frac{\sqrt{2}}{3}D_{u}\left[\tr\hat{\ep}\,-3\ep_{zz}\right]\\
	-\sqrt{\frac{2}{3}}\left[D_{u}(\ep_{xx}-\ep_{yy})+2D'_{u}i\ep_{xy}\right] & \sqrt{\frac{2}{3}}D'_{u}(\ep_{zx}+i\ep_{yz})
	\end{pmatrix}.
\end{align}
\end{widetext}
The Hamiltonian $\hat H(\kb,\hat\ep)$ respects the symmetry of the system. The terms containing the tensor components $\ep_{ij}$ and those containing the products $k_ik_j$ are located at the same places in the matrices $\hat H_i\cdot\hat\ep$ and $\hat H\0(\kb)$, respectively.

The values of the deformation potential parameters in Si are $D_{u}\approx3.3$ eV and $D'_{u}\approx4.42$ eV, while in GaAs $D_{u}\approx3.0$ eV and $D'_{u}\approx4.67$ eV. 

One can easily check that the eigenvalues of the upper left $4\times4$ block of $\hat{H}\0(\kb)+\hat{H}_{i}\cdot\hat{\ep}$ give the dispersion laws of the light and heavy holes in the presence of strain, Eq. (\ref{eq:strained_dispersion}). The parameters in Eq. (\ref{eq:strained_dispersion}) are $A=(\hbar^2/2m_{0})\gamma_{1}$, $B=(\hbar^2/2m_{0})2\gamma_{2}$, $C=(\hbar^2/2m_{0})\sqrt{12(\gamma_{3}^2-\gamma_{2}^2)}$, $D=(\hbar^2/2m_{0})2\sqrt{3}\gamma_{3}$, $a=-D_{d}$, $b=2D_{u}/3$, and $d=2D'_{u}/\sqrt{3}$.

A relation similar to Eq. (\ref{eq:E1_symmetry_two_band}), i.e.,
\begin{align}
	\label{eq:E1_symmetry_three_band}
	\int d^3\kb \left[E_\nu\1(\kb,\hat\ep)-a\tr\hat{\ep}\, \right]f(E_\nu\0,\mu\0)=0,
\end{align}
holds for the three-band model as well. This relation was used, in particular, to simplify the calculation of $\Delta\mathcal{F}\4$. Equation (\ref{eq:E1_symmetry_three_band}) follows from simple symmetry arguments. The integral over $\kb$ of the term $\propto E_\nu\1$ is an invariant. Therefore, it is proportional to $\tr\hat\ep$. To find the proportionality coefficient we can evaluate $E_\nu\1$ for hydrostatic strain $\ep_{ij} =\ep_0\delta_{ij}$. As seen from Eqs.~(\ref{eq:full_perturbation}) and (\ref{eq:off_diagonal_H_i}), such strain shifts the hole energies in all bands by $-3D_d \ep_0$. Then $E_\nu\1(\kb,\hat\ep) = -D_d\tr \hat \ep = a\tr\hat\ep$, which proves Eq.~(\ref{eq:E1_symmetry_three_band}).

Another useful relation that we have used in the main text refers to the higher-order corrections to the hole energy $E_\nu\un$. Since $\tr \hat H(\kb,\hat\ep) =2\sum_\nu E_\nu(\kb,\hat\ep)$, we have 
\begin{align}
\label{eq:trace_simple}
\sum_\nu E_\nu\un(\kb,\hat\ep) = 0 \quad \mathrm{for} \quad n>1.
\end{align}


\section{Temperature dependence of $c_{44}$}
\label{sec:c44_T_depen}

To better understand the nonmonotonic temperature dependence of $\Delta_\ka^{(\mathrm{h})}$ shown in Fig.~\ref{fig:Si_three_band_linear}, here we present separately the temperature dependence of the intrinsic and doping-induced parts of the elasticity parameter $c_{44}$. The value of $\sqrt{c_{44}}$ determines the eigenfrequency of the $\braket{110}$ Lam\'e mode, see Eq.~(\ref{eq:mode_elasticity}). As shown in Fig.~\ref{fig:Si_c44_T_depen}~(a) and (b), $c_{44}$ of pure Si monotonically decreases, whereas the doping-induced correction $c_{44}^{(\text{h})}$ monotonically increases with the increasing $T$. For a nondegenerate hole gas, this can be related to the estimate of the coefficient in the $n$th-order expansion term of the correction to the free energy $\lVert\widehat\Lambda^{(n)}\rVert\propto \mathcal{D}(\mathcal{D}/E_\text{kin})^{n-1}$, cf. Eq. (\ref{eq:delta_F_series}). It indicates that $c_{44}^{(\text{h})}\propto-\mathcal{D}^{2}/E_{\text{kin}} \propto -T^{-1}$; the negative sign here is due to the fact that the strain-induced redistribution of the holes always tends to lower the free energy. 

For the density $\nh=2.0\times10^{19}$ cm$^{-3}$ used in Fig.~\ref{fig:Si_c44_T_depen}, the hole gas is moderately degenerate in the light and heavy-hole bands and nondegenerate in the split-off band. Therefore, one still may think that $E_{\mathrm{kin}}$ goes up with the increasing temperature, leading to the decrease of $|c_{44}^{(\text{h})}|$. The competition between the opposite temperature dependence of the intrinsic and doping-induced contributions leads to the nonmonotonic behavior of $c_{44}$ shown by the lower dashed curve in Fig. \ref{fig:Si_c44_T_depen}~(a). In turn, this results in the nonmonotonic temperature dependence of $\Delta_\ka^{(\mathrm{h})}$ for the $\braket{110}$ Lam\'e mode.

The analysis of the doping-induced corrections to the second-order elastic constants for other modes is similar. However, the calculation shows that $|c_{ij}^{(\text{h})}|$ decreases with the increasing temperature even for the hole density $1.0\times10^{20}$~cm$^{-3}$. This differs from the nonmonotonic dependence on temperature of the doping-induced part of the fourth-order elastic constant $c_{ijkl}^{(\text{h})}$, cf. Fig. \ref{fig:Si_three_band_nonlinear_T_depen}~(b). The difference stems from the different structures of the expressions (\ref{eq:F_second}) and (\ref{eq:F_fourth}) for the quadratic and quartic corrections in $\hat\ep$ to the free energy.

%

\begin{thebibliography}{69}%
\makeatletter
\providecommand \@ifxundefined [1]{%
 \@ifx{#1\undefined}
}%
\providecommand \@ifnum [1]{%
 \ifnum #1\expandafter \@firstoftwo
 \else \expandafter \@secondoftwo
 \fi
}%
\providecommand \@ifx [1]{%
 \ifx #1\expandafter \@firstoftwo
 \else \expandafter \@secondoftwo
 \fi
}%
\providecommand \natexlab [1]{#1}%
\providecommand \enquote  [1]{``#1''}%
\providecommand \bibnamefont  [1]{#1}%
\providecommand \bibfnamefont [1]{#1}%
\providecommand \citenamefont [1]{#1}%
\providecommand \href@noop [0]{\@secondoftwo}%
\providecommand \href [0]{\begingroup \@sanitize@url \@href}%
\providecommand \@href[1]{\@@startlink{#1}\@@href}%
\providecommand \@@href[1]{\endgroup#1\@@endlink}%
\providecommand \@sanitize@url [0]{\catcode `\\12\catcode `\$12\catcode
  `\&12\catcode `\#12\catcode `\^12\catcode `\_12\catcode `\%12\relax}%
\providecommand \@@startlink[1]{}%
\providecommand \@@endlink[0]{}%
\providecommand \url  [0]{\begingroup\@sanitize@url \@url }%
\providecommand \@url [1]{\endgroup\@href {#1}{\urlprefix }}%
\providecommand \urlprefix  [0]{URL }%
\providecommand \Eprint [0]{\href }%
\providecommand \doibase [0]{https://doi.org/}%
\providecommand \selectlanguage [0]{\@gobble}%
\providecommand \bibinfo  [0]{\@secondoftwo}%
\providecommand \bibfield  [0]{\@secondoftwo}%
\providecommand \translation [1]{[#1]}%
\providecommand \BibitemOpen [0]{}%
\providecommand \bibitemStop [0]{}%
\providecommand \bibitemNoStop [0]{.\EOS\space}%
\providecommand \EOS [0]{\spacefactor3000\relax}%
\providecommand \BibitemShut  [1]{\csname bibitem#1\endcsname}%
\let\auto@bib@innerbib\@empty
\bibitem [{\citenamefont {Keyes}(1961)}]{Keyes1961}%
  \BibitemOpen
  \bibfield  {author} {\bibinfo {author} {\bibfnamefont {R.~W.}\ \bibnamefont
  {Keyes}},\ }\bibfield  {title} {\bibinfo {title} {The {{Electronic
  Contribution}} to the {{Elastic Properties}} of {{Germanium}}},\ }\href
  {https://doi.org/10.1147/rd.54.0266} {\bibfield  {journal} {\bibinfo
  {journal} {IBM J. Res. Dev.}\ }\textbf {\bibinfo {volume} {5}},\ \bibinfo
  {pages} {266} (\bibinfo {year} {1961})}\BibitemShut {NoStop}%
\bibitem [{\citenamefont {Keyes}()}]{Keyes1968}%
  \BibitemOpen
  \bibfield  {author} {\bibinfo {author} {\bibfnamefont {R.~W.}\ \bibnamefont
  {Keyes}},\ }\bibfield  {title} {\bibinfo {title} {Electronic effects in the
  elastic properties of semiconductors},\ }in\ \href
  {https://doi.org/10.1016/S0081-1947(08)60217-9} {\emph {\bibinfo {booktitle}
  {Solid State Physics}}},\ Vol.~\bibinfo {volume} {20},\ \bibinfo {editor}
  {edited by\ \bibinfo {editor} {\bibfnamefont {F.}~\bibnamefont {Seitz}},
  \bibinfo {editor} {\bibfnamefont {D.}~\bibnamefont {Turnbull}},\ and\
  \bibinfo {editor} {\bibfnamefont {H.}~\bibnamefont {Ehrenreich}}}\ (\bibinfo
  {publisher} {Academic Press})\ pp.\ \bibinfo {pages} {37--90}\BibitemShut
  {NoStop}%
\bibitem [{\citenamefont {Bir}\ and\ \citenamefont {Tursunov}(1963)}]{Bir1963}%
  \BibitemOpen
  \bibfield  {author} {\bibinfo {author} {\bibfnamefont {G.~L.}\ \bibnamefont
  {Bir}}\ and\ \bibinfo {author} {\bibfnamefont {A.}~\bibnamefont {Tursunov}},\
  }\bibfield  {title} {\bibinfo {title} {Effect of holes on the elastic
  constants of {{Germanium}}},\ }\href@noop {} {\bibfield  {journal} {\bibinfo
  {journal} {Sov. Phys. Solid State}\ }\textbf {\bibinfo {volume} {4}},\
  \bibinfo {pages} {1925} (\bibinfo {year} {1963})}\BibitemShut {NoStop}%
\bibitem [{\citenamefont {Cerdeira}\ and\ \citenamefont
  {Cardona}(1972)}]{Cerdeira1972}%
  \BibitemOpen
  \bibfield  {author} {\bibinfo {author} {\bibfnamefont {F.}~\bibnamefont
  {Cerdeira}}\ and\ \bibinfo {author} {\bibfnamefont {M.}~\bibnamefont
  {Cardona}},\ }\bibfield  {title} {\bibinfo {title} {Effect of {{Carrier
  Concentration}} on the {{Raman Frequencies}} of {{Si}} and {{Ge}}},\ }\href
  {https://doi.org/10.1103/PhysRevB.5.1440} {\bibfield  {journal} {\bibinfo
  {journal} {Phys. Rev. B}\ }\textbf {\bibinfo {volume} {5}},\ \bibinfo {pages}
  {1440} (\bibinfo {year} {1972})}\BibitemShut {NoStop}%
\bibitem [{\citenamefont {Kim}\ \emph {et~al.}(1976)\citenamefont {Kim},
  \citenamefont {Cardona},\ and\ \citenamefont {Rodriguez}}]{Kim1976}%
  \BibitemOpen
  \bibfield  {author} {\bibinfo {author} {\bibfnamefont {C.~K.}\ \bibnamefont
  {Kim}}, \bibinfo {author} {\bibfnamefont {M.}~\bibnamefont {Cardona}},\ and\
  \bibinfo {author} {\bibfnamefont {S.}~\bibnamefont {Rodriguez}},\ }\bibfield
  {title} {\bibinfo {title} {Effect of free carriers on the elastic constants
  of $p$-type silicon and germanium},\ }\href
  {https://doi.org/10.1103/PhysRevB.13.5429} {\bibfield  {journal} {\bibinfo
  {journal} {Phys. Rev. B}\ }\textbf {\bibinfo {volume} {13}},\ \bibinfo
  {pages} {5429} (\bibinfo {year} {1976})}\BibitemShut {NoStop}%
\bibitem [{\citenamefont {Averkiev}\ \emph {et~al.}(1984)\citenamefont
  {Averkiev}, \citenamefont {Ilisavskiy},\ and\ \citenamefont
  {Sternin}}]{Averkiev1984}%
  \BibitemOpen
  \bibfield  {author} {\bibinfo {author} {\bibfnamefont {N.~S.}\ \bibnamefont
  {Averkiev}}, \bibinfo {author} {\bibfnamefont {{\relax Yu}.~V.}\ \bibnamefont
  {Ilisavskiy}},\ and\ \bibinfo {author} {\bibfnamefont {V.~M.}\ \bibnamefont
  {Sternin}},\ }\bibfield  {title} {\bibinfo {title} {The free charge carrier
  effects on elastic properties of silicon},\ }\href
  {https://doi.org/10.1016/0038-1098(84)90709-9} {\bibfield  {journal}
  {\bibinfo  {journal} {Solid State Communications}\ }\textbf {\bibinfo
  {volume} {52}},\ \bibinfo {pages} {17} (\bibinfo {year} {1984})}\BibitemShut
  {NoStop}%
\bibitem [{\citenamefont {Khan}\ and\ \citenamefont {Allen}(1985)}]{Khan1985}%
  \BibitemOpen
  \bibfield  {author} {\bibinfo {author} {\bibfnamefont {F.~S.}\ \bibnamefont
  {Khan}}\ and\ \bibinfo {author} {\bibfnamefont {P.~B.}\ \bibnamefont
  {Allen}},\ }\bibfield  {title} {\bibinfo {title} {Temperature {{Dependence}}
  of the {{Elastic Constants}} of p$^{+}$ {{Silicon}}},\ }\href
  {https://doi.org/10.1002/pssb.2221280104} {\bibfield  {journal} {\bibinfo
  {journal} {Phys. Status Solidi B}\ }\textbf {\bibinfo {volume} {128}},\
  \bibinfo {pages} {31} (\bibinfo {year} {1985})}\BibitemShut {NoStop}%
\bibitem [{\citenamefont {Averkiev}\ \emph {et~al.}(1987)\citenamefont
  {Averkiev}, \citenamefont {Ilisavskiy},\ and\ \citenamefont
  {Sternin}}]{Averkiev1987}%
  \BibitemOpen
  \bibfield  {author} {\bibinfo {author} {\bibfnamefont {N.~S.}\ \bibnamefont
  {Averkiev}}, \bibinfo {author} {\bibfnamefont {{\relax Yu}.~V.}\ \bibnamefont
  {Ilisavskiy}},\ and\ \bibinfo {author} {\bibfnamefont {V.~M.}\ \bibnamefont
  {Sternin}},\ }\bibfield  {title} {\bibinfo {title} {Electron anharmonicity in
  semiconductors with a complex valence band structure},\ }\href@noop {}
  {\bibfield  {journal} {\bibinfo  {journal} {Sov. Phys. Solid State}\ }\textbf
  {\bibinfo {volume} {28}},\ \bibinfo {pages} {829} (\bibinfo {year}
  {1987})}\BibitemShut {NoStop}%
\bibitem [{\citenamefont {Cleland}\ and\ \citenamefont
  {Roukes}(1996)}]{Cleland1996}%
  \BibitemOpen
  \bibfield  {author} {\bibinfo {author} {\bibfnamefont {A.~N.}\ \bibnamefont
  {Cleland}}\ and\ \bibinfo {author} {\bibfnamefont {M.~L.}\ \bibnamefont
  {Roukes}},\ }\bibfield  {title} {\bibinfo {title} {Fabrication of {{High
  Frequency Nanometer Scale Mechanical Resonators}} from {{Bulk Si
  Crystals}}},\ }\href@noop {} {\bibfield  {journal} {\bibinfo  {journal}
  {Appl. Phys. Lett.}\ }\textbf {\bibinfo {volume} {69}},\ \bibinfo {pages}
  {2653} (\bibinfo {year} {1996})}\BibitemShut {NoStop}%
\bibitem [{\citenamefont {Cleland}\ and\ \citenamefont
  {Roukes}(1998)}]{Cleland1998}%
  \BibitemOpen
  \bibfield  {author} {\bibinfo {author} {\bibfnamefont {A.~N.}\ \bibnamefont
  {Cleland}}\ and\ \bibinfo {author} {\bibfnamefont {M.~L.}\ \bibnamefont
  {Roukes}},\ }\bibfield  {title} {\bibinfo {title} {A {{Nanometre-Scale
  Mechanical Electrometer}}},\ }\href@noop {} {\bibfield  {journal} {\bibinfo
  {journal} {Nature}\ }\textbf {\bibinfo {volume} {392}},\ \bibinfo {pages}
  {160} (\bibinfo {year} {1998})}\BibitemShut {NoStop}%
\bibitem [{\citenamefont {{van Beek}}\ and\ \citenamefont
  {Puers}(2011)}]{Beek2011}%
  \BibitemOpen
  \bibfield  {author} {\bibinfo {author} {\bibfnamefont {J.}~\bibnamefont {{van
  Beek}}}\ and\ \bibinfo {author} {\bibfnamefont {R.}~\bibnamefont {Puers}},\
  }\bibfield  {title} {\bibinfo {title} {A {{Review}} of {{MEMS Oscillators}}
  for {{Frequency Reference}} and {{Timing Applications}}},\ }\href@noop {}
  {\bibfield  {journal} {\bibinfo  {journal} {J. Micromech. Microeng.}\
  }\textbf {\bibinfo {volume} {22}},\ \bibinfo {pages} {013001} (\bibinfo
  {year} {2011})}\BibitemShut {NoStop}%
\bibitem [{\citenamefont {Yamaguchi}(2017)}]{Yamaguchi2017}%
  \BibitemOpen
  \bibfield  {author} {\bibinfo {author} {\bibfnamefont {H.}~\bibnamefont
  {Yamaguchi}},\ }\bibfield  {title} {\bibinfo {title} {{{GaAs-Based
  Micro}}/{{Nanomechanical Resonators}}},\ }\href@noop {} {\bibfield  {journal}
  {\bibinfo  {journal} {Semicond. Sci. Techn.}\ }\textbf {\bibinfo {volume}
  {32}},\ \bibinfo {pages} {103003} (\bibinfo {year} {2017})}\BibitemShut
  {NoStop}%
\bibitem [{\citenamefont {Miller}\ \emph {et~al.}(2018)\citenamefont {Miller},
  \citenamefont {Ansari}, \citenamefont {Heinz}, \citenamefont {Chen},
  \citenamefont {Flader}, \citenamefont {Shin}, \citenamefont {Villanueva},\
  and\ \citenamefont {Kenny}}]{Miller2018}%
  \BibitemOpen
  \bibfield  {author} {\bibinfo {author} {\bibfnamefont {J.~M.~L.}\
  \bibnamefont {Miller}}, \bibinfo {author} {\bibfnamefont {A.}~\bibnamefont
  {Ansari}}, \bibinfo {author} {\bibfnamefont {D.~B.}\ \bibnamefont {Heinz}},
  \bibinfo {author} {\bibfnamefont {Y.}~\bibnamefont {Chen}}, \bibinfo {author}
  {\bibfnamefont {I.~B.}\ \bibnamefont {Flader}}, \bibinfo {author}
  {\bibfnamefont {D.~D.}\ \bibnamefont {Shin}}, \bibinfo {author}
  {\bibfnamefont {L.~G.}\ \bibnamefont {Villanueva}},\ and\ \bibinfo {author}
  {\bibfnamefont {T.~W.}\ \bibnamefont {Kenny}},\ }\bibfield  {title} {\bibinfo
  {title} {Effective quality factor tuning mechanisms in micromechanical
  resonators},\ }\href {https://doi.org/10.1063/1.5027850} {\bibfield
  {journal} {\bibinfo  {journal} {Applied Physics Reviews}\ }\textbf {\bibinfo
  {volume} {5}},\ \bibinfo {pages} {041307} (\bibinfo {year}
  {2018})}\BibitemShut {NoStop}%
\bibitem [{\citenamefont {Bachtold}\ \emph {et~al.}(2022)\citenamefont
  {Bachtold}, \citenamefont {Moser},\ and\ \citenamefont
  {Dykman}}]{Bachtold2022a}%
  \BibitemOpen
  \bibfield  {author} {\bibinfo {author} {\bibfnamefont {A.}~\bibnamefont
  {Bachtold}}, \bibinfo {author} {\bibfnamefont {J.}~\bibnamefont {Moser}},\
  and\ \bibinfo {author} {\bibfnamefont {M.~I.}\ \bibnamefont {Dykman}},\
  }\bibfield  {title} {\bibinfo {title} {Mesoscopic physics of nanomechanical
  systems},\ }\href {https://doi.org/10.1103/RevModPhys.94.045005} {\bibfield
  {journal} {\bibinfo  {journal} {Rev. Mod. Phys.}\ }\textbf {\bibinfo {volume}
  {94}},\ \bibinfo {pages} {045005} (\bibinfo {year} {2022})}\BibitemShut
  {NoStop}%
\bibitem [{\citenamefont {Ng}\ \emph {et~al.}(2013)\citenamefont {Ng},
  \citenamefont {Lee}, \citenamefont {Ahn}, \citenamefont {Melamud},\ and\
  \citenamefont {Kenny}}]{Ng2013}%
  \BibitemOpen
  \bibfield  {author} {\bibinfo {author} {\bibfnamefont {E.~J.}\ \bibnamefont
  {Ng}}, \bibinfo {author} {\bibfnamefont {H.~K.}\ \bibnamefont {Lee}},
  \bibinfo {author} {\bibfnamefont {C.~H.}\ \bibnamefont {Ahn}}, \bibinfo
  {author} {\bibfnamefont {R.}~\bibnamefont {Melamud}},\ and\ \bibinfo {author}
  {\bibfnamefont {T.~W.}\ \bibnamefont {Kenny}},\ }\bibfield  {title} {\bibinfo
  {title} {Stability of {{Silicon Microelectromechanical Systems Resonant
  Thermometers}}},\ }\href {https://doi.org/10.1109/JSEN.2012.2227708}
  {\bibfield  {journal} {\bibinfo  {journal} {IEEE Sens. J.}\ }\textbf
  {\bibinfo {volume} {13}},\ \bibinfo {pages} {987} (\bibinfo {year}
  {2013})}\BibitemShut {NoStop}%
\bibitem [{\citenamefont {Shahmohammadi}\ \emph {et~al.}(2013)\citenamefont
  {Shahmohammadi}, \citenamefont {Fatemi},\ and\ \citenamefont
  {Abdolvand}}]{Shahmohammadi2013}%
  \BibitemOpen
  \bibfield  {author} {\bibinfo {author} {\bibfnamefont {M.}~\bibnamefont
  {Shahmohammadi}}, \bibinfo {author} {\bibfnamefont {H.}~\bibnamefont
  {Fatemi}},\ and\ \bibinfo {author} {\bibfnamefont {R.}~\bibnamefont
  {Abdolvand}},\ }\bibfield  {title} {\bibinfo {title} {Nonlinearity reduction
  in silicon resonators by doping and re-orientation},\ }in\ \href
  {https://doi.org/10.1109/MEMSYS.2013.6474362} {\emph {\bibinfo {booktitle}
  {2013 IEEE 26th International Conference on Micro Electro Mechanical Systems
  (MEMS)}}}\ (\bibinfo  {publisher} {IEEE},\ \bibinfo {year} {2013})\ pp.\
  \bibinfo {pages} {793--796}\BibitemShut {NoStop}%
\bibitem [{\citenamefont {Yang}\ \emph {et~al.}(2016)\citenamefont {Yang},
  \citenamefont {Ng}, \citenamefont {Polunin}, \citenamefont {Chen},
  \citenamefont {Flader}, \citenamefont {Shaw}, \citenamefont {Dykman},\ and\
  \citenamefont {Kenny}}]{Yang2016}%
  \BibitemOpen
  \bibfield  {author} {\bibinfo {author} {\bibfnamefont {Y.}~\bibnamefont
  {Yang}}, \bibinfo {author} {\bibfnamefont {E.~J.}\ \bibnamefont {Ng}},
  \bibinfo {author} {\bibfnamefont {P.~M.}\ \bibnamefont {Polunin}}, \bibinfo
  {author} {\bibfnamefont {Y.}~\bibnamefont {Chen}}, \bibinfo {author}
  {\bibfnamefont {I.~B.}\ \bibnamefont {Flader}}, \bibinfo {author}
  {\bibfnamefont {S.~W.}\ \bibnamefont {Shaw}}, \bibinfo {author}
  {\bibfnamefont {M.~I.}\ \bibnamefont {Dykman}},\ and\ \bibinfo {author}
  {\bibfnamefont {T.~W.}\ \bibnamefont {Kenny}},\ }\bibfield  {title} {\bibinfo
  {title} {Nonlinearity of {{Degenerately Doped Bulk-Mode Silicon MEMS
  Resonators}}},\ }\href@noop {} {\bibfield  {journal} {\bibinfo  {journal}
  {JMEMS}\ }\textbf {\bibinfo {volume} {25}},\ \bibinfo {pages} {859} (\bibinfo
  {year} {2016})}\BibitemShut {NoStop}%
\bibitem [{\citenamefont {Khazaeili}\ and\ \citenamefont
  {Abdolvand}(2020)}]{Khazaeili2020}%
  \BibitemOpen
  \bibfield  {author} {\bibinfo {author} {\bibfnamefont {B.}~\bibnamefont
  {Khazaeili}}\ and\ \bibinfo {author} {\bibfnamefont {R.}~\bibnamefont
  {Abdolvand}},\ }\bibfield  {title} {\bibinfo {title} {Inadequacy of
  third-order elastic coefficients for predicting nonlinearity in highly
  n-type-doped silicon resonators},\ }\href
  {https://doi.org/10.1109/TED.2019.2961946} {\bibfield  {journal} {\bibinfo
  {journal} {IEEE Transactions on Electron Devices}\ }\textbf {\bibinfo
  {volume} {67}},\ \bibinfo {pages} {614} (\bibinfo {year} {2020})}\BibitemShut
  {NoStop}%
\bibitem [{\citenamefont {Moskovtsev}\ and\ \citenamefont
  {Dykman}(2017)}]{moskovtsev2017strong}%
  \BibitemOpen
  \bibfield  {author} {\bibinfo {author} {\bibfnamefont {K.}~\bibnamefont
  {Moskovtsev}}\ and\ \bibinfo {author} {\bibfnamefont {M.~I.}\ \bibnamefont
  {Dykman}},\ }\bibfield  {title} {\bibinfo {title} {Strong vibration
  nonlinearity in semiconductor-based nanomechanical systems},\ }\href
  {https://doi.org/10.1103/PhysRevB.95.085426} {\bibfield  {journal} {\bibinfo
  {journal} {Phys. Rev. B}\ }\textbf {\bibinfo {volume} {95}},\ \bibinfo
  {pages} {085426} (\bibinfo {year} {2017})}\BibitemShut {NoStop}%
\bibitem [{\citenamefont {Bir}\ and\ \citenamefont {Pikus}(1974)}]{Bir1974}%
  \BibitemOpen
  \bibfield  {author} {\bibinfo {author} {\bibfnamefont {G.~L.}\ \bibnamefont
  {Bir}}\ and\ \bibinfo {author} {\bibfnamefont {G.~E.}\ \bibnamefont
  {Pikus}},\ }\href@noop {} {\emph {\bibinfo {title} {Symmetry and
  {{Strain-Induced Effects}} in {{Semiconductors}}}}}\ (\bibinfo  {publisher}
  {Wiley, N.-Y.},\ \bibinfo {year} {1974})\BibitemShut {NoStop}%
\bibitem [{\citenamefont {Hensel}\ \emph {et~al.}(1965)\citenamefont {Hensel},
  \citenamefont {Hasegawa},\ and\ \citenamefont {Nakayama}}]{Hensel1965}%
  \BibitemOpen
  \bibfield  {author} {\bibinfo {author} {\bibfnamefont {J.~C.}\ \bibnamefont
  {Hensel}}, \bibinfo {author} {\bibfnamefont {H.}~\bibnamefont {Hasegawa}},\
  and\ \bibinfo {author} {\bibfnamefont {M.}~\bibnamefont {Nakayama}},\
  }\bibfield  {title} {\bibinfo {title} {Cyclotron {{Resonance}} in
  {{Uniaxially Stressed Silicon}}. {{II}}. {{Nature}} of the {{Covalent
  Bond}}},\ }\href@noop {} {\bibfield  {journal} {\bibinfo  {journal} {Phys.
  Rev.}\ }\textbf {\bibinfo {volume} {138}},\ \bibinfo {pages} {A225} (\bibinfo
  {year} {1965})}\BibitemShut {NoStop}%
\bibitem [{\citenamefont {Laude}\ \emph {et~al.}(1971)\citenamefont {Laude},
  \citenamefont {Pollak},\ and\ \citenamefont {Cardona}}]{Laude1971}%
  \BibitemOpen
  \bibfield  {author} {\bibinfo {author} {\bibfnamefont {L.~D.}\ \bibnamefont
  {Laude}}, \bibinfo {author} {\bibfnamefont {F.~H.}\ \bibnamefont {Pollak}},\
  and\ \bibinfo {author} {\bibfnamefont {M.}~\bibnamefont {Cardona}},\
  }\bibfield  {title} {\bibinfo {title} {Effects of {{Uniaxial Stress}} on the
  {{Indirect Exciton Spectrum}} of {{Silicon}}},\ }\href@noop {} {\bibfield
  {journal} {\bibinfo  {journal} {Phys. Rev. B}\ }\textbf {\bibinfo {volume}
  {3}},\ \bibinfo {pages} {2623} (\bibinfo {year} {1971})}\BibitemShut
  {NoStop}%
\bibitem [{\citenamefont {Akhiezer}(1938)}]{Akhiezer1938}%
  \BibitemOpen
  \bibfield  {author} {\bibinfo {author} {\bibfnamefont {A.~I.}\ \bibnamefont
  {Akhiezer}},\ }\bibfield  {title} {\bibinfo {title} {On the {{Sound
  Absorption}} in {{Solids}}},\ }\href@noop {} {\bibfield  {journal} {\bibinfo
  {journal} {Zh. Eksp. Teor. Fiz.}\ }\textbf {\bibinfo {volume} {8}},\ \bibinfo
  {pages} {1318} (\bibinfo {year} {1938})}\BibitemShut {NoStop}%
\bibitem [{\citenamefont {Iyer}\ and\ \citenamefont
  {Candler}(2016)}]{Iyer2016}%
  \BibitemOpen
  \bibfield  {author} {\bibinfo {author} {\bibfnamefont {S.~S.}\ \bibnamefont
  {Iyer}}\ and\ \bibinfo {author} {\bibfnamefont {R.~N.}\ \bibnamefont
  {Candler}},\ }\bibfield  {title} {\bibinfo {title} {Mode- and
  {{Direction-Dependent Mechanical Energy Dissipation}} in {{Single-Crystal
  Resonators Due}} to {{Anharmonic Phonon-Phonon Scattering}}},\ }\href
  {https://doi.org/10.1103/PhysRevApplied.5.034002} {\bibfield  {journal}
  {\bibinfo  {journal} {Phys Rev Appl.}\ }\textbf {\bibinfo {volume} {5}},\
  \bibinfo {pages} {034002} (\bibinfo {year} {2016})}\BibitemShut {NoStop}%
\bibitem [{\citenamefont {Rodriguez}\ \emph {et~al.}(2019)\citenamefont
  {Rodriguez}, \citenamefont {Chandorkar}, \citenamefont {Watson},
  \citenamefont {Glaze}, \citenamefont {Ahn}, \citenamefont {Ng}, \citenamefont
  {Yang},\ and\ \citenamefont {Kenny}}]{Rodriguez2019}%
  \BibitemOpen
  \bibfield  {author} {\bibinfo {author} {\bibfnamefont {J.}~\bibnamefont
  {Rodriguez}}, \bibinfo {author} {\bibfnamefont {S.~A.}\ \bibnamefont
  {Chandorkar}}, \bibinfo {author} {\bibfnamefont {C.~A.}\ \bibnamefont
  {Watson}}, \bibinfo {author} {\bibfnamefont {G.~M.}\ \bibnamefont {Glaze}},
  \bibinfo {author} {\bibfnamefont {C.~H.}\ \bibnamefont {Ahn}}, \bibinfo
  {author} {\bibfnamefont {E.~J.}\ \bibnamefont {Ng}}, \bibinfo {author}
  {\bibfnamefont {Y.}~\bibnamefont {Yang}},\ and\ \bibinfo {author}
  {\bibfnamefont {T.~W.}\ \bibnamefont {Kenny}},\ }\bibfield  {title} {\bibinfo
  {title} {Direct {{Detection}} of {{Akhiezer Damping}} in a {{Silicon MEMS
  Resonator}}},\ }\href {https://doi.org/10.1038/s41598-019-38847-6} {\bibfield
   {journal} {\bibinfo  {journal} {Sci. Rep.}\ }\textbf {\bibinfo {volume}
  {9}},\ \bibinfo {pages} {2244} (\bibinfo {year} {2019})}\BibitemShut
  {NoStop}%
\bibitem [{\citenamefont {Terrazos}\ \emph {et~al.}(2021)\citenamefont
  {Terrazos}, \citenamefont {Marcellina}, \citenamefont {Wang}, \citenamefont
  {Coppersmith}, \citenamefont {Friesen}, \citenamefont {Hamilton},
  \citenamefont {Hu}, \citenamefont {Koiller}, \citenamefont {Saraiva},
  \citenamefont {Culcer},\ and\ \citenamefont {Capaz}}]{Terrazos2021}%
  \BibitemOpen
  \bibfield  {author} {\bibinfo {author} {\bibfnamefont {L.~A.}\ \bibnamefont
  {Terrazos}}, \bibinfo {author} {\bibfnamefont {E.}~\bibnamefont
  {Marcellina}}, \bibinfo {author} {\bibfnamefont {Z.}~\bibnamefont {Wang}},
  \bibinfo {author} {\bibfnamefont {S.~N.}\ \bibnamefont {Coppersmith}},
  \bibinfo {author} {\bibfnamefont {M.}~\bibnamefont {Friesen}}, \bibinfo
  {author} {\bibfnamefont {A.~R.}\ \bibnamefont {Hamilton}}, \bibinfo {author}
  {\bibfnamefont {X.}~\bibnamefont {Hu}}, \bibinfo {author} {\bibfnamefont
  {B.}~\bibnamefont {Koiller}}, \bibinfo {author} {\bibfnamefont {A.~L.}\
  \bibnamefont {Saraiva}}, \bibinfo {author} {\bibfnamefont {D.}~\bibnamefont
  {Culcer}},\ and\ \bibinfo {author} {\bibfnamefont {R.~B.}\ \bibnamefont
  {Capaz}},\ }\bibfield  {title} {\bibinfo {title} {Theory of hole-spin qubits
  in strained germanium quantum dots},\ }\href
  {https://doi.org/10.1103/PhysRevB.103.125201} {\bibfield  {journal} {\bibinfo
   {journal} {Phys. Rev. B}\ }\textbf {\bibinfo {volume} {103}},\ \bibinfo
  {pages} {125201} (\bibinfo {year} {2021})},\ \Eprint
  {https://arxiv.org/abs/1803.10320} {arXiv:1803.10320 [cond-mat]} \BibitemShut
  {NoStop}%
\bibitem [{\citenamefont {Landau}\ and\ \citenamefont
  {Lifshitz}(1997)}]{Landau1997}%
  \BibitemOpen
  \bibfield  {author} {\bibinfo {author} {\bibfnamefont {L.~D.}\ \bibnamefont
  {Landau}}\ and\ \bibinfo {author} {\bibfnamefont {E.~M.}\ \bibnamefont
  {Lifshitz}},\ }\href@noop {} {\emph {\bibinfo {title} {Quantum {{Mechanics}}.
  {{Non-Relativistic Theory}}}}},\ \bibinfo {edition} {3rd}\ ed.\ (\bibinfo
  {publisher} {Butterworth-Heinemann, Oxford},\ \bibinfo {year}
  {1997})\BibitemShut {NoStop}%
\bibitem [{\citenamefont {Nye}(2002)}]{Nye2002}%
  \BibitemOpen
  \bibfield  {author} {\bibinfo {author} {\bibfnamefont {J.~F.}\ \bibnamefont
  {Nye}},\ }\href@noop {} {\emph {\bibinfo {title} {Physical {{Properties Of
  Crystals}}: {{Their Representation}} by {{Tensors}} and {{Matrices}}}}}\
  (\bibinfo  {publisher} {Oxford University Press, U.S.A.},\ \bibinfo {address}
  {Oxford Oxfordshire : New York},\ \bibinfo {year} {2002})\BibitemShut
  {NoStop}%
\bibitem [{\citenamefont {Landau}\ and\ \citenamefont
  {Lifshitz}(1986)}]{Landau1986}%
  \BibitemOpen
  \bibfield  {author} {\bibinfo {author} {\bibfnamefont {L.}~\bibnamefont
  {Landau}}\ and\ \bibinfo {author} {\bibfnamefont {E.}~\bibnamefont
  {Lifshitz}},\ }\href@noop {} {\emph {\bibinfo {title} {Theory of
  {{Elasticity}}}}},\ \bibinfo {edition} {3rd}\ ed.\ (\bibinfo  {publisher}
  {Butterworth-Heinemann Ltd., Oxford},\ \bibinfo {year} {1986})\BibitemShut
  {NoStop}%
\bibitem [{\citenamefont {Seeger}(2004)}]{Seeger2004}%
  \BibitemOpen
  \bibfield  {author} {\bibinfo {author} {\bibfnamefont {K.}~\bibnamefont
  {Seeger}},\ }\href@noop {} {\emph {\bibinfo {title} {Semiconductor
  {{Physics}}: {{An Introduction}}}}},\ \bibinfo {edition} {9th}\ ed.\
  (\bibinfo  {publisher} {Springer-Verlag, Berlin},\ \bibinfo {year}
  {2004})\BibitemShut {NoStop}%
\bibitem [{\citenamefont {Jordan}(1980)}]{jordan1980evaluation}%
  \BibitemOpen
  \bibfield  {author} {\bibinfo {author} {\bibfnamefont {A.}~\bibnamefont
  {Jordan}},\ }\bibfield  {title} {\bibinfo {title} {An evaluation of the
  thermal and elastic constants affecting {GaAs} crystal growth},\ }\href
  {https://doi.org/https://doi.org/10.1016/0022-0248(80)90287-0} {\bibfield
  {journal} {\bibinfo  {journal} {Journal of Crystal Growth}\ }\textbf
  {\bibinfo {volume} {49}},\ \bibinfo {pages} {631} (\bibinfo {year}
  {1980})}\BibitemShut {NoStop}%
\bibitem [{\citenamefont {Ng}\ \emph {et~al.}(2015)\citenamefont {Ng},
  \citenamefont {Hong}, \citenamefont {Yang}, \citenamefont {Ahn},
  \citenamefont {Evenhart},\ and\ \citenamefont {Kenny}}]{Ng2015}%
  \BibitemOpen
  \bibfield  {author} {\bibinfo {author} {\bibfnamefont {E.~J.}\ \bibnamefont
  {Ng}}, \bibinfo {author} {\bibfnamefont {V.}~\bibnamefont {Hong}}, \bibinfo
  {author} {\bibfnamefont {Y.}~\bibnamefont {Yang}}, \bibinfo {author}
  {\bibfnamefont {{\relax Ch}.~H.}\ \bibnamefont {Ahn}}, \bibinfo {author}
  {\bibfnamefont {C.~L.~M.}\ \bibnamefont {Evenhart}},\ and\ \bibinfo {author}
  {\bibfnamefont {T.~W.}\ \bibnamefont {Kenny}},\ }\bibfield  {title} {\bibinfo
  {title} {Temperature {{Dependence}} of the {{Elastic Constants}} of {{Doped
  Silicon}}},\ }\href@noop {} {\bibfield  {journal} {\bibinfo  {journal}
  {JMEMS}\ }\textbf {\bibinfo {volume} {24}},\ \bibinfo {pages} {730} (\bibinfo
  {year} {2015})}\BibitemShut {NoStop}%
\bibitem [{\citenamefont {Lee}(2016)}]{Lee2016}%
  \BibitemOpen
  \bibfield  {author} {\bibinfo {author} {\bibfnamefont {J.~E.-Y.}\
  \bibnamefont {Lee}},\ }\bibfield  {title} {\bibinfo {title} {Lam{\'e} {{Mode
  MEMS Resonators}}},\ }in\ \href
  {https://doi.org/10.1007/978-94-007-6178-0_101001-1} {\emph {\bibinfo
  {booktitle} {Encyclopedia of {{Nanotechnology}}}}},\ \bibinfo {editor}
  {edited by\ \bibinfo {editor} {\bibfnamefont {B.}~\bibnamefont {Bhushan}}}\
  (\bibinfo  {publisher} {Springer Netherlands},\ \bibinfo {address}
  {Dordrecht},\ \bibinfo {year} {2016})\ pp.\ \bibinfo {pages}
  {1--9}\BibitemShut {NoStop}%
\bibitem [{\citenamefont {Graff}(1991)}]{graff1991wave}%
  \BibitemOpen
  \bibfield  {author} {\bibinfo {author} {\bibfnamefont {K.~F.}\ \bibnamefont
  {Graff}},\ }\href@noop {} {\emph {\bibinfo {title} {Wave Motion in Elastic
  Solids}}}\ (\bibinfo  {publisher} {Dover, New York},\ \bibinfo {year}
  {1991})\BibitemShut {NoStop}%
\bibitem [{\citenamefont {Hamoumi}\ \emph {et~al.}(2018)\citenamefont
  {Hamoumi}, \citenamefont {Allain}, \citenamefont {Hease}, \citenamefont
  {Gil-Santos}, \citenamefont {Morgenroth}, \citenamefont {G\'erard},
  \citenamefont {Lema\^{\i}tre}, \citenamefont {Leo},\ and\ \citenamefont
  {Favero}}]{Hamoumi2018}%
  \BibitemOpen
  \bibfield  {author} {\bibinfo {author} {\bibfnamefont {M.}~\bibnamefont
  {Hamoumi}}, \bibinfo {author} {\bibfnamefont {P.~E.}\ \bibnamefont {Allain}},
  \bibinfo {author} {\bibfnamefont {W.}~\bibnamefont {Hease}}, \bibinfo
  {author} {\bibfnamefont {E.}~\bibnamefont {Gil-Santos}}, \bibinfo {author}
  {\bibfnamefont {L.}~\bibnamefont {Morgenroth}}, \bibinfo {author}
  {\bibfnamefont {B.}~\bibnamefont {G\'erard}}, \bibinfo {author}
  {\bibfnamefont {A.}~\bibnamefont {Lema\^{\i}tre}}, \bibinfo {author}
  {\bibfnamefont {G.}~\bibnamefont {Leo}},\ and\ \bibinfo {author}
  {\bibfnamefont {I.}~\bibnamefont {Favero}},\ }\bibfield  {title} {\bibinfo
  {title} {Microscopic {{Nanomechanical Dissipation}} in {{Gallium Arsenide
  Resonators}}},\ }\href {https://doi.org/10.1103/PhysRevLett.120.223601}
  {\bibfield  {journal} {\bibinfo  {journal} {Phys. Rev. Lett.}\ }\textbf
  {\bibinfo {volume} {120}},\ \bibinfo {pages} {223601} (\bibinfo {year}
  {2018})}\BibitemShut {NoStop}%
\bibitem [{\citenamefont {Allain}\ \emph {et~al.}(2021)\citenamefont {Allain},
  \citenamefont {Guha}, \citenamefont {Baker}, \citenamefont {Parrain},
  \citenamefont {Lema{\^i}tre}, \citenamefont {Leo},\ and\ \citenamefont
  {Favero}}]{Allain2021}%
  \BibitemOpen
  \bibfield  {author} {\bibinfo {author} {\bibfnamefont {P.~E.}\ \bibnamefont
  {Allain}}, \bibinfo {author} {\bibfnamefont {B.}~\bibnamefont {Guha}},
  \bibinfo {author} {\bibfnamefont {C.}~\bibnamefont {Baker}}, \bibinfo
  {author} {\bibfnamefont {D.}~\bibnamefont {Parrain}}, \bibinfo {author}
  {\bibfnamefont {A.}~\bibnamefont {Lema{\^i}tre}}, \bibinfo {author}
  {\bibfnamefont {G.}~\bibnamefont {Leo}},\ and\ \bibinfo {author}
  {\bibfnamefont {I.}~\bibnamefont {Favero}},\ }\bibfield  {title} {\bibinfo
  {title} {Electro-optomechanical modulation instability in a semiconductor
  resonator},\ }\href {https://doi.org/10.1103/PhysRevLett.126.243901}
  {\bibfield  {journal} {\bibinfo  {journal} {Phys. Rev. Lett.}\ }\textbf
  {\bibinfo {volume} {126}},\ \bibinfo {pages} {243901} (\bibinfo {year}
  {2021})}\BibitemShut {NoStop}%
\bibitem [{\citenamefont {Varshni}(1970)}]{varshni1970temperature}%
  \BibitemOpen
  \bibfield  {author} {\bibinfo {author} {\bibfnamefont {Y.~P.}\ \bibnamefont
  {Varshni}},\ }\bibfield  {title} {\bibinfo {title} {Temperature dependence of
  the elastic constants},\ }\href {https://doi.org/10.1103/PhysRevB.2.3952}
  {\bibfield  {journal} {\bibinfo  {journal} {Phys. Rev. B}\ }\textbf {\bibinfo
  {volume} {2}},\ \bibinfo {pages} {3952} (\bibinfo {year} {1970})}\BibitemShut
  {NoStop}%
\bibitem [{\citenamefont {Winkler}(2003)}]{winkler2003spin}%
  \BibitemOpen
  \bibfield  {author} {\bibinfo {author} {\bibfnamefont {R.}~\bibnamefont
  {Winkler}},\ }\href@noop {} {\emph {\bibinfo {title} {Spin-orbit Coupling
  Effects in Two-Dimensional Electron and Hole Systems}}}\ (\bibinfo
  {publisher} {Springer Berlin, Heidelberg},\ \bibinfo {year}
  {2003})\BibitemShut {NoStop}%
\bibitem [{\citenamefont {Burenkov}\ \emph {et~al.}(1973)\citenamefont
  {Burenkov}, \citenamefont {Burdukov}, \citenamefont {Davidov},\ and\
  \citenamefont {Nikaronov}}]{burenkov1973temprature}%
  \BibitemOpen
  \bibfield  {author} {\bibinfo {author} {\bibfnamefont {Y.~A.}\ \bibnamefont
  {Burenkov}}, \bibinfo {author} {\bibfnamefont {Y.~M.}\ \bibnamefont
  {Burdukov}}, \bibinfo {author} {\bibfnamefont {S.~Y.}\ \bibnamefont
  {Davidov}},\ and\ \bibinfo {author} {\bibfnamefont {S.~P.}\ \bibnamefont
  {Nikaronov}},\ }\bibfield  {title} {\bibinfo {title} {Temperature dependences
  of the elastic constants of gallium arsenide},\ }\href@noop {} {\bibfield
  {journal} {\bibinfo  {journal} {Sov. Phys. Solid State}\ }\textbf {\bibinfo
  {volume} {15}},\ \bibinfo {pages} {1175} (\bibinfo {year}
  {1973})}\BibitemShut {NoStop}%
\bibitem [{\citenamefont {Cottam}\ and\ \citenamefont
  {Saunders}(1973)}]{cottam1973elastic}%
  \BibitemOpen
  \bibfield  {author} {\bibinfo {author} {\bibfnamefont {R.~I.}\ \bibnamefont
  {Cottam}}\ and\ \bibinfo {author} {\bibfnamefont {G.~A.}\ \bibnamefont
  {Saunders}},\ }\bibfield  {title} {\bibinfo {title} {The elastic constants of
  {GaAs} from 2 {K} to 320 {K}},\ }\href
  {https://doi.org/10.1088/0022-3719/6/13/011} {\bibfield  {journal} {\bibinfo
  {journal} {Journal of Physics C: Solid State Physics}\ }\textbf {\bibinfo
  {volume} {6}},\ \bibinfo {pages} {2105} (\bibinfo {year} {1973})}\BibitemShut
  {NoStop}%
\bibitem [{\citenamefont {Landau}\ and\ \citenamefont
  {Lifshitz}(2004)}]{Landau2004a}%
  \BibitemOpen
  \bibfield  {author} {\bibinfo {author} {\bibfnamefont {L.~D.}\ \bibnamefont
  {Landau}}\ and\ \bibinfo {author} {\bibfnamefont {E.~M.}\ \bibnamefont
  {Lifshitz}},\ }\href@noop {} {\emph {\bibinfo {title} {Mechanics}}},\
  \bibinfo {edition} {3rd}\ ed.\ (\bibinfo  {publisher} {Elsevier, Amsterdam},\
  \bibinfo {year} {2004})\BibitemShut {NoStop}%
\bibitem [{\citenamefont {Dykman}\ and\ \citenamefont
  {Krivoglaz}(1984)}]{Dykman1984}%
  \BibitemOpen
  \bibfield  {author} {\bibinfo {author} {\bibfnamefont {M.~I.}\ \bibnamefont
  {Dykman}}\ and\ \bibinfo {author} {\bibfnamefont {M.~A.}\ \bibnamefont
  {Krivoglaz}},\ }\bibfield  {title} {\bibinfo {title} {Theory of {{Nonlinear
  Oscillators Interacting}} with a {{Medium}}},\ }in\ \href@noop {} {\emph
  {\bibinfo {booktitle} {Sov. {{Phys}}. {{Reviews}}}}},\ Vol.~\bibinfo {volume}
  {5},\ \bibinfo {editor} {edited by\ \bibinfo {editor} {\bibfnamefont {I.~M.}\
  \bibnamefont {Khalatnikov}}}\ (\bibinfo  {publisher} {Harwood Academic, New
  York},\ \bibinfo {year} {1984})\ pp.\ \bibinfo {pages} {265--441,
  \url{https://web.pa.msu.edu/people/dykman/pub06/DKreview84.pdf}}\BibitemShut
  {NoStop}%
\bibitem [{\citenamefont {Barnard}\ \emph {et~al.}(2012)\citenamefont
  {Barnard}, \citenamefont {Sazonova}, \citenamefont {{van der Zande}},\ and\
  \citenamefont {McEuen}}]{Barnard2012}%
  \BibitemOpen
  \bibfield  {author} {\bibinfo {author} {\bibfnamefont {A.~W.}\ \bibnamefont
  {Barnard}}, \bibinfo {author} {\bibfnamefont {V.}~\bibnamefont {Sazonova}},
  \bibinfo {author} {\bibfnamefont {A.~M.}\ \bibnamefont {{van der Zande}}},\
  and\ \bibinfo {author} {\bibfnamefont {P.~L.}\ \bibnamefont {McEuen}},\
  }\bibfield  {title} {\bibinfo {title} {Fluctuation {{Broadening}} in {{Carbon
  Nanotube Resonators}}},\ }\href@noop {} {\bibfield  {journal} {\bibinfo
  {journal} {PNAS}\ }\textbf {\bibinfo {volume} {109}},\ \bibinfo {pages}
  {19093} (\bibinfo {year} {2012})}\BibitemShut {NoStop}%
\bibitem [{\citenamefont {Venstra}\ \emph {et~al.}(2012)\citenamefont
  {Venstra}, \citenamefont {{van Leeuwen}},\ and\ \citenamefont {{van der
  Zant}}}]{Venstra2012}%
  \BibitemOpen
  \bibfield  {author} {\bibinfo {author} {\bibfnamefont {W.~J.}\ \bibnamefont
  {Venstra}}, \bibinfo {author} {\bibfnamefont {R.}~\bibnamefont {{van
  Leeuwen}}},\ and\ \bibinfo {author} {\bibfnamefont {H.~S.~J.}\ \bibnamefont
  {{van der Zant}}},\ }\bibfield  {title} {\bibinfo {title} {Strongly {{Coupled
  Modes}} in a {{Weakly Driven Micromechanical Resonator}}},\ }\href@noop {}
  {\bibfield  {journal} {\bibinfo  {journal} {Appl. Phys. Lett.}\ }\textbf
  {\bibinfo {volume} {101}},\ \bibinfo {pages} {243111} (\bibinfo {year}
  {2012})}\BibitemShut {NoStop}%
\bibitem [{\citenamefont {Matheny}\ \emph {et~al.}(2013)\citenamefont
  {Matheny}, \citenamefont {Villanueva}, \citenamefont {Karabalin},
  \citenamefont {Sader},\ and\ \citenamefont {Roukes}}]{Matheny2013}%
  \BibitemOpen
  \bibfield  {author} {\bibinfo {author} {\bibfnamefont {M.~H.}\ \bibnamefont
  {Matheny}}, \bibinfo {author} {\bibfnamefont {L.~G.}\ \bibnamefont
  {Villanueva}}, \bibinfo {author} {\bibfnamefont {R.~B.}\ \bibnamefont
  {Karabalin}}, \bibinfo {author} {\bibfnamefont {J.~E.}\ \bibnamefont
  {Sader}},\ and\ \bibinfo {author} {\bibfnamefont {M.~L.}\ \bibnamefont
  {Roukes}},\ }\bibfield  {title} {\bibinfo {title} {Nonlinear
  {{Mode-Coupling}} in {{Nanomechanical Systems}}},\ }\href@noop {} {\bibfield
  {journal} {\bibinfo  {journal} {Nano Lett.}\ }\textbf {\bibinfo {volume}
  {13}},\ \bibinfo {pages} {1622} (\bibinfo {year} {2013})}\BibitemShut
  {NoStop}%
\bibitem [{\citenamefont {Gieseler}\ \emph {et~al.}(2013)\citenamefont
  {Gieseler}, \citenamefont {Novotny},\ and\ \citenamefont
  {Quidant}}]{Gieseler2013}%
  \BibitemOpen
  \bibfield  {author} {\bibinfo {author} {\bibfnamefont {J.}~\bibnamefont
  {Gieseler}}, \bibinfo {author} {\bibfnamefont {L.}~\bibnamefont {Novotny}},\
  and\ \bibinfo {author} {\bibfnamefont {R.}~\bibnamefont {Quidant}},\
  }\bibfield  {title} {\bibinfo {title} {Thermal {{Nonlinearities}} in a
  {{Nanomechanical Oscillator}}},\ }\href@noop {} {\bibfield  {journal}
  {\bibinfo  {journal} {Nat. Phys.}\ }\textbf {\bibinfo {volume} {9}},\
  \bibinfo {pages} {806} (\bibinfo {year} {2013})}\BibitemShut {NoStop}%
\bibitem [{\citenamefont {Miao}\ \emph {et~al.}(2014)\citenamefont {Miao},
  \citenamefont {Yeom}, \citenamefont {Wang}, \citenamefont {Standley},\ and\
  \citenamefont {Bockrath}}]{Miao2014}%
  \BibitemOpen
  \bibfield  {author} {\bibinfo {author} {\bibfnamefont {T.~F.}\ \bibnamefont
  {Miao}}, \bibinfo {author} {\bibfnamefont {S.}~\bibnamefont {Yeom}}, \bibinfo
  {author} {\bibfnamefont {P.}~\bibnamefont {Wang}}, \bibinfo {author}
  {\bibfnamefont {B.}~\bibnamefont {Standley}},\ and\ \bibinfo {author}
  {\bibfnamefont {M.}~\bibnamefont {Bockrath}},\ }\bibfield  {title} {\bibinfo
  {title} {Graphene {{Nanoelectromechanical Systems}} as {{Stochastic-Frequency
  Oscillators}}},\ }\href {https://doi.org/10.1021/nl403936a} {\bibfield
  {journal} {\bibinfo  {journal} {Nano Lett.}\ }\textbf {\bibinfo {volume}
  {14}},\ \bibinfo {pages} {2982} (\bibinfo {year} {2014})}\BibitemShut
  {NoStop}%
\bibitem [{\citenamefont {Maillet}\ \emph {et~al.}(2017)\citenamefont
  {Maillet}, \citenamefont {Zhou}, \citenamefont {Gazizulin}, \citenamefont
  {Maldonado~Cid}, \citenamefont {Defoort}, \citenamefont {Fefferman},
  \citenamefont {Bourgeois},\ and\ \citenamefont {Collin}}]{Maillet2017}%
  \BibitemOpen
  \bibfield  {author} {\bibinfo {author} {\bibfnamefont {O.}~\bibnamefont
  {Maillet}}, \bibinfo {author} {\bibfnamefont {X.}~\bibnamefont {Zhou}},
  \bibinfo {author} {\bibfnamefont {R.}~\bibnamefont {Gazizulin}}, \bibinfo
  {author} {\bibfnamefont {A.}~\bibnamefont {Maldonado~Cid}}, \bibinfo {author}
  {\bibfnamefont {M.}~\bibnamefont {Defoort}}, \bibinfo {author} {\bibfnamefont
  {A.~D.}\ \bibnamefont {Fefferman}}, \bibinfo {author} {\bibfnamefont
  {O.}~\bibnamefont {Bourgeois}},\ and\ \bibinfo {author} {\bibfnamefont
  {E.}~\bibnamefont {Collin}},\ }\bibfield  {title} {\bibinfo {title}
  {Non-{{Linear Frequency Transduction}} of {{Nano-Mechanical Brownian
  Motion}}},\ }\href@noop {} {\bibfield  {journal} {\bibinfo  {journal} {Phys.
  Rev. B}\ }\textbf {\bibinfo {volume} {96}},\ \bibinfo {pages} {165434}
  (\bibinfo {year} {2017})}\BibitemShut {NoStop}%
\bibitem [{\citenamefont {Huang}\ \emph {et~al.}(2019)\citenamefont {Huang},
  \citenamefont {Soskin}, \citenamefont {Khovanov}, \citenamefont {Mannella},
  \citenamefont {Ninios},\ and\ \citenamefont {Chan}}]{Huang2019}%
  \BibitemOpen
  \bibfield  {author} {\bibinfo {author} {\bibfnamefont {L.}~\bibnamefont
  {Huang}}, \bibinfo {author} {\bibfnamefont {S.~M.}\ \bibnamefont {Soskin}},
  \bibinfo {author} {\bibfnamefont {I.~A.}\ \bibnamefont {Khovanov}}, \bibinfo
  {author} {\bibfnamefont {R.}~\bibnamefont {Mannella}}, \bibinfo {author}
  {\bibfnamefont {K.}~\bibnamefont {Ninios}},\ and\ \bibinfo {author}
  {\bibfnamefont {H.~B.}\ \bibnamefont {Chan}},\ }\bibfield  {title} {\bibinfo
  {title} {Frequency stabilization and noise-induced spectral narrowing in
  resonators with zero dispersion},\ }\href@noop {} {\bibfield  {journal}
  {\bibinfo  {journal} {Nat. Commun.}\ }\textbf {\bibinfo {volume} {10}},\
  \bibinfo {pages} {3930} (\bibinfo {year} {2019})}\BibitemShut {NoStop}%
\bibitem [{\citenamefont {Amarouchene}\ \emph {et~al.}(2019)\citenamefont
  {Amarouchene}, \citenamefont {Mangeat}, \citenamefont {Montes}, \citenamefont
  {Ondic}, \citenamefont {Gu{\'e}rin}, \citenamefont {Dean},\ and\
  \citenamefont {Louyer}}]{Amarouchene2019}%
  \BibitemOpen
  \bibfield  {author} {\bibinfo {author} {\bibfnamefont {Y.}~\bibnamefont
  {Amarouchene}}, \bibinfo {author} {\bibfnamefont {M.}~\bibnamefont
  {Mangeat}}, \bibinfo {author} {\bibfnamefont {B.~V.}\ \bibnamefont {Montes}},
  \bibinfo {author} {\bibfnamefont {L.}~\bibnamefont {Ondic}}, \bibinfo
  {author} {\bibfnamefont {T.}~\bibnamefont {Gu{\'e}rin}}, \bibinfo {author}
  {\bibfnamefont {D.~S.}\ \bibnamefont {Dean}},\ and\ \bibinfo {author}
  {\bibfnamefont {Y.}~\bibnamefont {Louyer}},\ }\bibfield  {title} {\bibinfo
  {title} {Nonequilibrium {{Dynamics Induced}} by {{Scattering Forces}} for
  {{Optically Trapped Nanoparticles}} in {{Strongly Inertial Regimes}}},\
  }\href {https://doi.org/10.1103/PhysRevLett.122.183901} {\bibfield  {journal}
  {\bibinfo  {journal} {Phys. Rev. Lett.}\ }\textbf {\bibinfo {volume} {122}},\
  \bibinfo {pages} {183901} (\bibinfo {year} {2019})}\BibitemShut {NoStop}%
\bibitem [{\citenamefont {Luttinger}\ and\ \citenamefont
  {Kohn}(1955)}]{Luttinger1955}%
  \BibitemOpen
  \bibfield  {author} {\bibinfo {author} {\bibfnamefont {J.~M.}\ \bibnamefont
  {Luttinger}}\ and\ \bibinfo {author} {\bibfnamefont {W.}~\bibnamefont
  {Kohn}},\ }\bibfield  {title} {\bibinfo {title} {Motion of electrons and
  holes in perturbed periodic fields},\ }\href
  {https://doi.org/10.1103/PhysRev.97.869} {\bibfield  {journal} {\bibinfo
  {journal} {Physical Review}\ }\textbf {\bibinfo {volume} {97}},\ \bibinfo
  {pages} {869} (\bibinfo {year} {1955})}\BibitemShut {NoStop}%
\bibitem [{\citenamefont {Fornberg}(1988)}]{fornberg1988generation}%
  \BibitemOpen
  \bibfield  {author} {\bibinfo {author} {\bibfnamefont {B.}~\bibnamefont
  {Fornberg}},\ }\bibfield  {title} {\bibinfo {title} {Generation of finite
  difference formulas on arbitrarily spaced grids},\ }\href
  {https://doi.org/10.1090%2FS0025-5718-1988-0935077-0} {\bibfield  {journal}
  {\bibinfo  {journal} {Mathematics of Computation}\ }\textbf {\bibinfo
  {volume} {51}},\ \bibinfo {pages} {699} (\bibinfo {year} {1988})}\BibitemShut
  {NoStop}%
\bibitem [{\citenamefont {Kittel}\ and\ \citenamefont
  {Mitchell}(1954)}]{kittel1954theory}%
  \BibitemOpen
  \bibfield  {author} {\bibinfo {author} {\bibfnamefont {C.}~\bibnamefont
  {Kittel}}\ and\ \bibinfo {author} {\bibfnamefont {A.~H.}\ \bibnamefont
  {Mitchell}},\ }\bibfield  {title} {\bibinfo {title} {Theory of donor and
  acceptor states in silicon and germanium},\ }\href
  {https://doi.org/10.1103/PhysRev.96.1488} {\bibfield  {journal} {\bibinfo
  {journal} {Phys. Rev.}\ }\textbf {\bibinfo {volume} {96}},\ \bibinfo {pages}
  {1488} (\bibinfo {year} {1954})}\BibitemShut {NoStop}%
\bibitem [{\citenamefont {Kittel}(2005)}]{kittel2005introduction}%
  \BibitemOpen
  \bibfield  {author} {\bibinfo {author} {\bibfnamefont {C.}~\bibnamefont
  {Kittel}},\ }\href@noop {} {\emph {\bibinfo {title} {Introduction to Solid
  State Physics}}}\ (\bibinfo  {publisher} {John Wiley \& Sons},\ \bibinfo
  {year} {2005})\BibitemShut {NoStop}%
\bibitem [{\citenamefont {Bourgeois}\ \emph {et~al.}(1997)\citenamefont
  {Bourgeois}, \citenamefont {Steinsland}, \citenamefont {Blanc},\ and\
  \citenamefont {{de Rooij}}}]{Bourgeois1997}%
  \BibitemOpen
  \bibfield  {author} {\bibinfo {author} {\bibfnamefont {C.}~\bibnamefont
  {Bourgeois}}, \bibinfo {author} {\bibfnamefont {E.}~\bibnamefont
  {Steinsland}}, \bibinfo {author} {\bibfnamefont {N.}~\bibnamefont {Blanc}},\
  and\ \bibinfo {author} {\bibfnamefont {N.}~\bibnamefont {{de Rooij}}},\
  }\bibfield  {title} {\bibinfo {title} {Design of resonators for the
  determination of the temperature coefficients of elastic constants of
  monocrystalline silicon},\ }in\ \href
  {https://doi.org/10.1109/FREQ.1997.639192} {\emph {\bibinfo {booktitle}
  {Proceedings of International Frequency Control Symposium}}}\ (\bibinfo
  {year} {1997})\ pp.\ \bibinfo {pages} {791--799}\BibitemShut {NoStop}%
\bibitem [{\citenamefont {Lyon}\ \emph {et~al.}(1977)\citenamefont {Lyon},
  \citenamefont {Salinger}, \citenamefont {Swenson},\ and\ \citenamefont
  {White}}]{Lyon1977}%
  \BibitemOpen
  \bibfield  {author} {\bibinfo {author} {\bibfnamefont {K.~G.}\ \bibnamefont
  {Lyon}}, \bibinfo {author} {\bibfnamefont {G.~L.}\ \bibnamefont {Salinger}},
  \bibinfo {author} {\bibfnamefont {C.~A.}\ \bibnamefont {Swenson}},\ and\
  \bibinfo {author} {\bibfnamefont {G.~K.}\ \bibnamefont {White}},\ }\bibfield
  {title} {\bibinfo {title} {Linear thermal expansion measurements on silicon
  from 6 to 340 k},\ }\href {https://doi.org/10.1063/1.323747} {\bibfield
  {journal} {\bibinfo  {journal} {Journal of Applied Physics}\ }\textbf
  {\bibinfo {volume} {48}},\ \bibinfo {pages} {865} (\bibinfo {year}
  {1977})}\BibitemShut {NoStop}%
\bibitem [{\citenamefont {Hall}(1967)}]{Hall1967}%
  \BibitemOpen
  \bibfield  {author} {\bibinfo {author} {\bibfnamefont {J.~J.}\ \bibnamefont
  {Hall}},\ }\bibfield  {title} {\bibinfo {title} {Electronic {{Effects}} in
  the {{Elastic Constants}} of $n$-{{Type Silicon}}},\ }\href
  {https://doi.org/10.1103/PhysRev.161.756} {\bibfield  {journal} {\bibinfo
  {journal} {Phys. Rev.}\ }\textbf {\bibinfo {volume} {161}},\ \bibinfo {pages}
  {756} (\bibinfo {year} {1967})}\BibitemShut {NoStop}%
\bibitem [{\citenamefont {Philip}\ and\ \citenamefont
  {Breazeale}(1983)}]{Philip1983}%
  \BibitemOpen
  \bibfield  {author} {\bibinfo {author} {\bibfnamefont {J.}~\bibnamefont
  {Philip}}\ and\ \bibinfo {author} {\bibfnamefont {M.~A.}\ \bibnamefont
  {Breazeale}},\ }\bibfield  {title} {\bibinfo {title} {Third-order elastic
  constants and {G}r{\"u}neisen parameters of silicon and germanium between 3
  and 300\,{$^\circ$}{K}},\ }\href {https://doi.org/10.1063/1.332033}
  {\bibfield  {journal} {\bibinfo  {journal} {Journal of Applied Physics}\
  }\textbf {\bibinfo {volume} {54}},\ \bibinfo {pages} {752} (\bibinfo {year}
  {1983})}\BibitemShut {NoStop}%
\bibitem [{\citenamefont {Regner}\ \emph {et~al.}(2013)\citenamefont {Regner},
  \citenamefont {Sellan}, \citenamefont {Su}, \citenamefont {Amon},
  \citenamefont {McGaughey},\ and\ \citenamefont {Malen}}]{Regner2013}%
  \BibitemOpen
  \bibfield  {author} {\bibinfo {author} {\bibfnamefont {K.~T.}\ \bibnamefont
  {Regner}}, \bibinfo {author} {\bibfnamefont {D.~P.}\ \bibnamefont {Sellan}},
  \bibinfo {author} {\bibfnamefont {Z.}~\bibnamefont {Su}}, \bibinfo {author}
  {\bibfnamefont {C.~H.}\ \bibnamefont {Amon}}, \bibinfo {author}
  {\bibfnamefont {A.~J.~H.}\ \bibnamefont {McGaughey}},\ and\ \bibinfo {author}
  {\bibfnamefont {J.~A.}\ \bibnamefont {Malen}},\ }\bibfield  {title} {\bibinfo
  {title} {Broadband phonon mean free path contributions to thermal
  conductivity measured using frequency domain thermoreflectance},\ }\href
  {https://doi.org/10.1038/ncomms2630} {\bibfield  {journal} {\bibinfo
  {journal} {Nature Communications}\ }\textbf {\bibinfo {volume} {4}},\
  \bibinfo {pages} {1640} (\bibinfo {year} {2013})}\BibitemShut {NoStop}%
\bibitem [{\citenamefont {Lifshitz}\ and\ \citenamefont
  {Pitaevskii}(1981)}]{lifshitz1981physical}%
  \BibitemOpen
  \bibfield  {author} {\bibinfo {author} {\bibfnamefont {E.~M.}\ \bibnamefont
  {Lifshitz}}\ and\ \bibinfo {author} {\bibfnamefont {L.~P.}\ \bibnamefont
  {Pitaevskii}},\ }\href@noop {} {\emph {\bibinfo {title} {Physical Kinetics:
  Course of Theoretical Physics, Volume 10 (First edition)}}}\ (\bibinfo
  {publisher} {Pergamon Press, Oxford},\ \bibinfo {year} {1981})\BibitemShut
  {NoStop}%
\bibitem [{\citenamefont {Gurevich}(1988)}]{Gurevich1988}%
  \BibitemOpen
  \bibfield  {author} {\bibinfo {author} {\bibfnamefont {V.~L.}\ \bibnamefont
  {Gurevich}},\ }\href@noop {} {\emph {\bibinfo {title} {Transport in {{Phonon
  Systems}}}}}\ (\bibinfo  {publisher} {Elsevier Science Ltd},\ \bibinfo {year}
  {1988})\BibitemShut {NoStop}%
\bibitem [{\citenamefont {Kiselev}\ and\ \citenamefont
  {Iafrate}(2008)}]{Kiselev2008}%
  \BibitemOpen
  \bibfield  {author} {\bibinfo {author} {\bibfnamefont {A.~A.}\ \bibnamefont
  {Kiselev}}\ and\ \bibinfo {author} {\bibfnamefont {G.~J.}\ \bibnamefont
  {Iafrate}},\ }\bibfield  {title} {\bibinfo {title} {Phonon {{Dynamics}} and
  {{Phonon Assisted Losses}} in {{Euler-Bernoulli Nanobeams}}},\ }\href
  {https://doi.org/10.1103/PhysRevB.77.205436} {\bibfield  {journal} {\bibinfo
  {journal} {Phys. Rev. B}\ }\textbf {\bibinfo {volume} {77}},\ \bibinfo
  {pages} {205436} (\bibinfo {year} {2008})}\BibitemShut {NoStop}%
\bibitem [{\citenamefont {Kunal}\ and\ \citenamefont
  {Aluru}(2014)}]{Kunal2014}%
  \BibitemOpen
  \bibfield  {author} {\bibinfo {author} {\bibfnamefont {K.}~\bibnamefont
  {Kunal}}\ and\ \bibinfo {author} {\bibfnamefont {N.~R.}\ \bibnamefont
  {Aluru}},\ }\bibfield  {title} {\bibinfo {title} {Intrinsic {{Dissipation}}
  in a {{Nano-Mechanical Resonator}}},\ }\href
  {https://doi.org/10.1063/1.4894282} {\bibfield  {journal} {\bibinfo
  {journal} {J. Appl. Phys.}\ }\textbf {\bibinfo {volume} {116}},\ \bibinfo
  {pages} {094304} (\bibinfo {year} {2014})}\BibitemShut {NoStop}%
\bibitem [{\citenamefont {Atalaya}\ \emph {et~al.}(2016)\citenamefont
  {Atalaya}, \citenamefont {Kenny}, \citenamefont {Roukes},\ and\ \citenamefont
  {Dykman}}]{Atalaya2016}%
  \BibitemOpen
  \bibfield  {author} {\bibinfo {author} {\bibfnamefont {J.}~\bibnamefont
  {Atalaya}}, \bibinfo {author} {\bibfnamefont {T.~W.}\ \bibnamefont {Kenny}},
  \bibinfo {author} {\bibfnamefont {M.~L.}\ \bibnamefont {Roukes}},\ and\
  \bibinfo {author} {\bibfnamefont {M.~I.}\ \bibnamefont {Dykman}},\ }\bibfield
   {title} {\bibinfo {title} {Nonlinear {{Damping}} and {{Dephasing}} in
  {{Nanomechanical Systems}}},\ }\href@noop {} {\bibfield  {journal} {\bibinfo
  {journal} {Phys. Rev. B}\ }\textbf {\bibinfo {volume} {94}},\ \bibinfo
  {pages} {195440} (\bibinfo {year} {2016})}\BibitemShut {NoStop}%
\bibitem [{\citenamefont {Woodruff}\ and\ \citenamefont
  {Ehrenreich}(1961)}]{Woodruff1961}%
  \BibitemOpen
  \bibfield  {author} {\bibinfo {author} {\bibfnamefont {T.~O.}\ \bibnamefont
  {Woodruff}}\ and\ \bibinfo {author} {\bibfnamefont {H.}~\bibnamefont
  {Ehrenreich}},\ }\bibfield  {title} {\bibinfo {title} {Absorption of {{Sound
  In Insulators}}},\ }\href@noop {} {\bibfield  {journal} {\bibinfo  {journal}
  {Phys. Rev.}\ }\textbf {\bibinfo {volume} {123}},\ \bibinfo {pages} {1553}
  (\bibinfo {year} {1961})}\BibitemShut {NoStop}%
\bibitem [{\citenamefont {Madelung}(2004)}]{madelung2004semiconductor}%
  \BibitemOpen
  \bibfield  {author} {\bibinfo {author} {\bibfnamefont {O.}~\bibnamefont
  {Madelung}},\ }\href@noop {} {\emph {\bibinfo {title} {Semiconductors: Data
  Handbook}}}\ (\bibinfo  {publisher} {Springer Berlin, Heidelberg},\ \bibinfo
  {year} {2004})\BibitemShut {NoStop}%
\bibitem [{\citenamefont {Gauster}(1971)}]{gauster1971lowtemp}%
  \BibitemOpen
  \bibfield  {author} {\bibinfo {author} {\bibfnamefont {W.~B.}\ \bibnamefont
  {Gauster}},\ }\bibfield  {title} {\bibinfo {title} {Low-temperature
  {G}r\"uneisen parameters for silicon and aluminum},\ }\href
  {https://doi.org/10.1103/PhysRevB.4.1288} {\bibfield  {journal} {\bibinfo
  {journal} {Phys. Rev. B}\ }\textbf {\bibinfo {volume} {4}},\ \bibinfo {pages}
  {1288} (\bibinfo {year} {1971})}\BibitemShut {NoStop}%
\bibitem [{\citenamefont {Asheghi}\ \emph {et~al.}(2002)\citenamefont
  {Asheghi}, \citenamefont {Kurabayashi}, \citenamefont {Kasnavi},\ and\
  \citenamefont {Goodson}}]{Asheghi2002}%
  \BibitemOpen
  \bibfield  {author} {\bibinfo {author} {\bibfnamefont {M.}~\bibnamefont
  {Asheghi}}, \bibinfo {author} {\bibfnamefont {K.}~\bibnamefont
  {Kurabayashi}}, \bibinfo {author} {\bibfnamefont {R.}~\bibnamefont
  {Kasnavi}},\ and\ \bibinfo {author} {\bibfnamefont {K.~E.}\ \bibnamefont
  {Goodson}},\ }\bibfield  {title} {\bibinfo {title} {Thermal conduction in
  doped single-crystal silicon films},\ }\href
  {https://doi.org/10.1063/1.1458057} {\bibfield  {journal} {\bibinfo
  {journal} {Journal of Applied Physics}\ }\textbf {\bibinfo {volume} {91}},\
  \bibinfo {pages} {5079} (\bibinfo {year} {2002})}\BibitemShut {NoStop}%
\bibitem [{\citenamefont {Dongre}\ \emph {et~al.}(2020)\citenamefont {Dongre},
  \citenamefont {Carrete}, \citenamefont {Wen}, \citenamefont {Ma},
  \citenamefont {Li}, \citenamefont {Mingo},\ and\ \citenamefont
  {Madsen}}]{dongre2020combined}%
  \BibitemOpen
  \bibfield  {author} {\bibinfo {author} {\bibfnamefont {B.}~\bibnamefont
  {Dongre}}, \bibinfo {author} {\bibfnamefont {J.}~\bibnamefont {Carrete}},
  \bibinfo {author} {\bibfnamefont {S.}~\bibnamefont {Wen}}, \bibinfo {author}
  {\bibfnamefont {J.}~\bibnamefont {Ma}}, \bibinfo {author} {\bibfnamefont
  {W.}~\bibnamefont {Li}}, \bibinfo {author} {\bibfnamefont {N.}~\bibnamefont
  {Mingo}},\ and\ \bibinfo {author} {\bibfnamefont {G.~K.~H.}\ \bibnamefont
  {Madsen}},\ }\bibfield  {title} {\bibinfo {title} {Combined treatment of
  phonon scattering by electrons and point defects explains the thermal
  conductivity reduction in highly-doped {Si}},\ }\href
  {https://doi.org/10.1039/C9TA11424F} {\bibfield  {journal} {\bibinfo
  {journal} {J. Mater. Chem. A}\ }\textbf {\bibinfo {volume} {8}},\ \bibinfo
  {pages} {1273} (\bibinfo {year} {2020})}\BibitemShut {NoStop}%
\end{thebibliography}

\end{document}